\documentclass[10pt,twocolumn,a4paper]{article}
\usepackage{adjustbox}
\usepackage{float}
\usepackage{amsmath,amsthm,amsfonts,amssymb,amscd,mathrsfs}
\usepackage{mathtools}
\usepackage{enumerate,titlesec,graphicx,hyperref}
\usepackage[utf8x]{inputenc}
\usepackage[italicdiff]{physics}
\usepackage{listings}
\usepackage{authblk}
\usepackage{appendix}
\usepackage{etoolbox}
\BeforeBeginEnvironment{appendices}{\clearpage}
\usepackage{lscape}
\usepackage{afterpage}
\usepackage{caption}
\usepackage{subcaption}
\usepackage{diagbox}
\usepackage[rightcaption]{sidecap}
\usepackage{xcolor}

\hypersetup{colorlinks=true,allcolors=blue}
\allowdisplaybreaks

\title{A Learned Closure Method Applied to Phase Mixing in a Turbulent Gradient-Driven Gyrokinetic System in Simple Geometry}
\author[1]{A. Shukla}
\author[1]{D. R. Hatch}
\author[2]{W. Dorland}
\author[1]{C. Michoski}

\date{}
\affil[1]{Institute for Fusion Studies, University of Texas at Austin}
\affil[2]{Department of Physics, University of Maryland} 

\begin{document}
\twocolumn[
  \begin{@twocolumnfalse}
    \maketitle
 \begin{abstract}
    
We present a new method for formulating closures that learn from kinetic simulation data. We apply this method to phase mixing in a simple gyrokinetic turbulent system---temperature gradient driven turbulence in an unsheared slab. The closure, called the learned multi-mode (LMM) closure, is constructed by, first, extracting an optimal basis from a nonlinear kinetic simulation using singular value decomposition (SVD). Subsequent nonlinear fluid simulations are projected onto this basis and the results are used to formulate the closure. We compare the closure with other closures schemes over a broad range of the relevant 2D parameter space (collisionality and gradient drive). We find that the turbulent kinetic system produces phase mixing rates much lower than the linear expectations, which the LMM closure is capable of capturing.  We also compare radial heat fluxes.  A Hammett-Perkins closure, generalized to include collisional effects, is quite successful throughout the parameter space, producing $\sim 14 \%$ Root-Mean-Square (RMS) error.  The LMM closure is also very effective: when trained at three (two) points (in a 35 point parameter grid), the LMM closure produces $9 \%$ ($12\%$) RMS errors.  The LMM procedure can be readily generalized to other closure problems.
    
\end{abstract}
  \end{@twocolumnfalse}
]

\section{Introduction}
\label{introduction}

%A kinetic description of a plasma defines the evolution of the distribution function for particles at position $\vec x$ and velocity $\vec v$ in time according to the Boltzmann Equation: %plasma kinetic equation:
%\begin{equation}
%\frac{\partial f_s}{\partial t}+\boldsymbol{v} \cdot \nabla %f_s+\frac{q}{m}\left(\boldsymbol{E}+\frac{\boldsymbol{v}}{c} %\times \boldsymbol{B}\right) \cdot \nabla_{v} f=C[f_s],
%\label{kinetic_equation}
%\end{equation}
%where $f_s(\boldsymbol{x},\boldsymbol{v})$ denotes the distribution function of particle species $s$, $\boldsymbol{E}$ is the electric field, $\boldsymbol{B}$ is the magnetic field, and C is a collision operator.  In order to calculate the self-consistent fields, this must be coupled to Maxwell's equations.

%The kinetic equation time-evolves a six dimensional phase space at the fast cyclotron frequency rendering it extremely challenging to solve numerically.  Although direct numerical simulations of Eq.~\ref{kinetic_equation} can be achieved at great expense for limited problems, the the full kinetic equation remains, perhaps, most valuable as a starting point for reduced treatments of plasmas.  

The gyrokinetic model~\cite{Freiman,krommes2012,abel_2013}, in which the fast gyration of particles around the magnetic field is averaged out, has proven to be a useful description of strongly magnetized plasmas. The kinetic system is reduced from 6D to 5D (3 spatial dimensions and 2 velocity dimensions) and the fast cyclotron timescale is eliminated.  The kinetic system is thus greatly simplified for both analysis and simulation.  %The gyrokinetic equation effectively evolves a distribution of `charged rings', and is expressed in terms of the guiding center coordinates and gyro-averaged fields. 

Gyrokinetics has become the standard tool for describing turbulent transport in magnetic fusion devices, and more broadly, has found fruitful applications ranging from basic plasma physics to space / astro systems~\cite{plunk_cowley_schekochihin_tatsuno_2010,mj2011,Told15,howes2008a}. In fusion applications, in particular, gyrokinetic simulations have demonstrated increasing explanatory power with respect to experimental observations~\cite{gorler14,Hatch_2016_microtearing,Holland16}. Despite these developments, nonlinear gyrokinetics remains too expensive to be routinely used to predict confinement (i.e., to evolve profiles) or broadly explore parameter space for optimal confinement configurations.  Consequently, further reductions in complexity remain highly desirable.  
    
One such approach to further reducing the gyrokinetic system, the gyrofluid framework, was introduced in Ref.~\cite{hammet90,dorland93}.  A critical component of gyrofluid models is closures that model important kinetic effects within a fluid treatment.  In this paper we study closures in a reduced gyrokinetic system with a Hermite polynomial basis (in velocity space) for a relatively simple turbulent system---gradient-driven turbulence in an unsheared slab.  The Hermite basis facilitates a direct comparison of closed fluid simulations (truncated at low Hermite number) with kinetic simulations (truncated at high Hermite number).   

A prototypical example of gyrofluid closures is the Hammett-Perkins (HP)\cite{hammet90} closure.  The HP procedure closes a fluid system using the linear kinetic response. This was a major breakthrough, providing a much more rigorous treatment of collisionless plasmas than conventional fluid theory.  It effectively models phase mixing / Landau damping rates resulting in linear growth rates and frequencies in quite good agreement with the true (kinetic) values.  Its utility is evidenced by continued vigorous development and application to broad-ranging plasma systems such as turbulent transport in tokamaks~\cite{staebler_2005,Staebler_2007}, tokamak edge turbulence~\cite{Scott_2007,xu_2013,Peer_2017}, and space plasma turbulence~\cite{hunana_2013,sulem_2015}.  In this paper, we examine the HP closure in a turbulent system.  

%However, its validity is perhaps not as well established in turbulent systems where nonlinearity can alter phase mixing dynamics.  As will be shown for the simple system studied here, the standard gyrofluid closures over-estimate the phase mixing rate in comparison with nonlinear kinetic simulations.  This is consistent with several recent papers that have noted that Landau damping rates can be greatly modified from linear expectations in turbulent systems (see, e.g. Refs.~\cite{plunk2013,kanekar_schekochihin_dorland_loureiro_2015,parker_2016,Hatch_2016,schekochihin_2016,Meyrand1185}).  Despite this limitation, the HP closure is quite effective in reproducing radial heat fluxes.  %\DRHnote{I need to describe the specific examples in detail here}  %\DRHnote{also add other closures approaches--Callen, Sugama, Mattor, Parker}
    
We also introduce a new method for learning closures from kinetic simulation data.  This method, which we call the learned multi-mode (LMM) closure, is motivated by the notion that a closure for a turbulent system may benefit from the versatility to capture aspects of the nonlinear state.  Our closure procedure first extracts, from a single nonlinear kinetic simulation, an optimal basis using singular value decomposition (SVD).  Subsequent fluid simulations are projected onto the `fluid' components of this basis and the projection is used to formulate the closure.  %and test it broadly throughout the 2D parameter space (gradient drive and collision frequency) of the system.  

The closure schemes are examined in nonlinear simulations over a broad range of parameter space through the lens of two metrics: (1) the phase mixing rate, and (2) the radial heat flux.  The HP closures substantially over-predict the phase mixing rates, which are greatly reduced in comparison with the linear predictions.  This is consistent with several recent papers that have noted that Landau damping rates can be greatly modified from linear expectations in turbulent systems (see, e.g. Refs.~\cite{plunk2013,kanekar_schekochihin_dorland_loureiro_2015,parker_2016,Hatch_2016,schekochihin_2016,Meyrand1185}).  In contrast, the LMM closure reproduces phase mixing rates quite accurately.    

Despite limitations in reproducing turbulent phase mixing rates, the HP closure is much more accurate in reproducing the kinetic values of the radial heat flux.  To be quantitative, an HP closure generalized to include collisional effects results in an RMS error of $~\sim 14 \%$ over the parameter space. 

The LMM closure produces accurate heat fluxes in regions near the training parameter point, with performance deteriorating with distance.  Training at multiple, sparsely separated, points results in a highly effective closure.  When trained at three (two) points (in a 35 point parameter grid), the LMM closure produces $9 \%$ ($12\%$) errors.  We envision the utility of the closure to be maximized within a rigorous statistical framework like Bayesian optimization to guide selection of training points. 

This paper is outlined as follows: in Sec.~\ref{reduced_hermite} we describe the simplified gyrokinetic model and DNA code, which we use to test the performance of various closures. In Sec.~\ref{closures}, we briefly describe HP-style closures and introduce the new learned multi-mode closure. In Sec.~\ref{closure_evaluation}, we analyze the linear and nonlinear validity of closures by comparing growth rates and phase mixing rates produced by each closure to those of the kinetic system. Sec.~\ref{simulation_results} evaluates the performance of each closure by comparing the radial heat fluxes throughout a broad parameter space. Advantages, limitations, and possible future avenues of research are described in the concluding section~\ref{summary}.

\section{Reduced Gyrokinetic Equations in a Hermite Representation}
\label{reduced_hermite}

In order to explore various closure ideas, we study a relatively simple kinetic turbulent system---gradient driven electrostatic instabilities and turbulence in an unsheared slab.  The underlying model is a reduction of gyrokinetics to one dimension (parallel to the magnetic field) in velocity space and retaining rudimentary finite Larmor radius (FLR) effects.  As a starting point, we consider the gyrokinetic equations for a single kinetic species $s$ in a triply periodic Fourier representation of an unsheared slab:

\begin{multline*}
\frac{\partial f_s}{\partial t} =-\left [\omega_n+\omega_T \left (v_{||}^2+\mu - \frac{3}{2} \right) \right ]F_{0s} i k_y \bar{\phi}_s \\ 
-\sqrt{2}v_{||} i k_z \left (f_s +  F_{0s} \bar{\phi}_s \right ) + C(f_s) \\
+ \sum_{\mathbf{k}^{\prime}_\perp}\left(k^{\prime}_x k_y - k_x k^{\prime}_y  \right )  \bar{\phi}_{s,\mathbf{k}^{\prime}_\perp}f_{s,\mathbf{k}_\perp-\mathbf{k}^{\prime}_\perp},
\label{gk_eq}
\end{multline*}
where the overbar denotes a gyroaverage (i.e. multiplication by the zeroth order Bessel function $J_0(\sqrt{2 \mu} k_\perp)$), $k_x (\rho_s)$ is the radial wavenumber (in the direction of the background gradients) and $\rho_s = \left ( \frac{T_s}{m_s} \right )^{1/2}\frac{m_s}{q_s B_0}$ is the gyroradius, $k_y (\rho_s)$ is the binormal (to the magnetic field and the $x$ direction) wavenumber, $k_\perp = \sqrt{k_x^2 + k_y^2}$, $k_z (L_{ref})$ is the parallel (to the magnetic field) wavenumber,
$v_{||} (1/ v_{th,s})$ is the parallel (to the magnetic field) velocity, the thermal velocity is defined as $v_{th,s}=\left ( \frac{2T_s}{m_s} \right )^{1/2}$, $\mu = \frac{m_s v_{\perp}^2}{2B_0}(B_0/T_{0s})$ is the magnetic moment and acts as the perpendicular velocity coordinate, $t (\frac{\sqrt{T_s/m_s}}{L_{ref}})$ is time, $C \rightarrow C (\frac{L_{ref}}{v_{th,s}})$ is a collision operator (defined below), $f_s (\frac{L_{ref}}{\rho_s}\frac{v_{th,s}^3}{n_{0s}})$ is the perturbed distribution function, $F_{0s} (\frac{v_{th,s}^3}{n_{0s}})$ is the background Maxwellian distribution function, $\phi (\frac{L_{ref}}{\rho_s}\frac{T_{0s}}{e})$ is the electrostatic potential,  $\omega_n=L_{ref}/L_n$ is the inverse normalized density gradient scale length, $\omega_T=L_{ref}/L_T$ is the inverse normalized temperature gradient scale length, and $L_{ref}$ is a reference macroscopic scale length (normalization is shown in parentheses).  The gyrocenter distribution function, $f_{k_x,k_y,k_z}(v_{||},\mu)$, is a function of the three spatial wavenumbers and two velocity coordinates.  %Note that the thermal velocity $v_{th,s}$ is defined  $v_{Ti} \rightarrow \sqrt{2}$. %\DRHnote{double check square roots of two in thermal velocity stuff}.   

%These equations are normalized as follows:
%\begin{align*}
%k_x \rightarrow k_x \rho_s \\
%k_y \rightarrow k_y \rho_s \\
%k_z \rightarrow k_z L_{ref} \\
%v_{||} \rightarrow v_{||} / v_{th,s}\\
%\mu \rightarrow \mu \frac{B_0}{T_{0s}} \\
%t \rightarrow t \frac{v_{th,s}}{L_{ref}} \\
%C \rightarrow C \frac{L_{ref}}{v_{th,s}} \\
%f_s \rightarrow f_s %\frac{L_{ref}}{\rho_s}\frac{v_{th,s}^3}{n_{0s}} \\
%F_{0s} \rightarrow F_{0s} \frac{v_{th,s}^3}{n_{0s}}\\
%\phi \rightarrow \phi \frac{L_{ref}}{\rho_s}\frac{T_{0s}}{e} 
%\end{align*}

The field equation for the electrostatic potential is,
\begin{equation}
\phi_{k_x,k_y}=\frac{ \int \bar{f}_s dv_{||} d\mu + \tau \langle \phi \rangle_{FS}\delta_{k_y,0}}{\tau + \left [1-\Gamma_0 (b) \right ] },
%\label{field_eqn}
\end{equation}
where $\tau$ is the ratio of ion to electron temperature, $\Gamma_0(x)=I_0(x)e^{-x}$ with $I_0(x)$ the zeroth order modified Bessel function, $b=k^2_{\perp}$, and the flux-surface averaged potential is, 
\begin{equation}
\langle \phi \rangle_{FS}=\frac{\pi \langle \int \bar{f_s} dv_{||} d\mu \rangle_{FS}}{\left[1- \Gamma_0(b) \right]}.
\label{field_eqn}
\end{equation}
The inclusion of the flux-surface averaged potential in Eq.~\ref{field_eqn} is appropriate for an ion species driven by the ion-temperature gradient (ITG). Such a system favors strong zonal flow production.  In this unsheared slab system, this results in strongly suppressed turbulence~\cite{Hatch13}.  Consequently, we neglect this term in our simulations, which is appropriate for an electron species with adiabatic ions.  The gradient drive is now due to the electron-temperature gradient (ETG).  In the following, the species labels are dropped with the understanding that all quantities should be considered to be electrons.  

The reduced system studied in this work is derived by first integrating over the $\mu$ coordinate and retaining rudimentary FLR effects of the form $J_0(\sqrt{2 \mu} k_\perp) \rightarrow e^{-k_\perp^2/2}$.  This approximation is exact only when integrating over a Maxwellian distribution function, as is done for all gyroaverages of the electrostatic potential. More sophisticated treatments are noted in the literature~\cite{dorland93}, but this rudimentary treatment is sufficient for our purposes, namely to stabilize the instabilities at $k_\perp \rho_s \gtrapprox 1$. 

The parallel velocity dimension is then decomposed on a basis of Hermite polynomials $f(v) = \sum_{n=0}^{\infty}  f_n H_n(v_{||}) e^{-v_{||}^2}$, where $n$ denotes now the order of the Hermite polynomial.  The Hermite representation facilitates analysis of the system in both fluid (truncation at low $n$) and kinetic (truncation at high $n$) limits and is thus well suited for studying closures.  There is a simple connection between these Hermite moments, $ f_n$, and the conventional fluid moments (see Appendix~\ref{hermite_appendix} for details).  The Hermite-based equations are as follows~\cite{Hatch13,jpp2014}:  

\begin{multline}
\pdv{ f_{\mathbf k, n}}{t} = \frac{\omega_T ik_y}{\pi^{1/4}}\frac{k^{2}_{\perp}}{2}\bar \phi_{\mathbf k} \delta_{n,0} 
            - \frac{\omega_n ik_y}{\pi^{1/4}} \bar \phi_{\mathbf k}\delta_{n,0} \\
            - \frac{\omega_T ik_y}{\sqrt 2 \pi^{1/4}} \bar \phi_{\mathbf k}\delta_{n,2}
            - \frac{ik_z}{\pi^{1/4}} \bar \phi_{\mathbf k}\delta_{n,1}\\
            -ik_z[\sqrt{n} f_{\mathbf k,n-1} + \sqrt{n+1} f_{\mathbf k, n+1}]
            -\nu n  f_{\mathbf k,n} \\
            +\sum_{\mathbf{k}'} (k'_x k_y - k_x k'_y)\bar\phi_{\mathbf{k}'}f_{\mathbf{k}-\mathbf{k}'}
\label{dnaeq}
\end{multline}

%with the following linear and nonlinear operators:
%\begin{equation}
%\begin{aligned}
%    L[f_n] =& \omega_T ik_y\frac{k^{2}_{\perp}}{2}e^{-k^{2}_{\perp}/2}\phi\delta_{n,0} 
            %- \omega_n ik_y e^{-k^{2}_{\perp}/2}\phi\delta_{n,0} \\
            %&- \omega_T ik_y e^{-k^{2}_{\perp}/2}\phi\delta_{n,2}
            %- ik_z e^{-k^{2}_{\perp}/2}\phi\delta_{n,1}\\
            %&-ik_z[\sqrt{n}g_{n-1} + \sqrt{n+1}g_{n+1}]
            %-\nu n g_n
%\end{aligned}
%\label{L}
%\end{equation}
%\begin{equation}
%N[g] = \sum_{\vec{k}'} (k'_x k_y - k_x k'_y)e^{-k'^{2}_{\perp}/2}\phi_{\vec{k}'}g_{\vec{k}-\vec{k}'}
%\label{N}
%\end{equation}
%where $\hat g_{\mathbf k,n}(\rho_i n_{i0}/L_nv_{ti}^3)$ is the ion distribution function, $n_{i0}$ is the ion density, $L_n$ is the density gradient length scale, $v_{ti}$ is the ion thermal velocity, $n$ denotes the order of the hermite polynomial, $t(L_n/v_{ti})$ is time, $\omega_n = L_{ref}/L_n$ gradient length scale, $\omega_T = L_{ref}/L_T$ is temperature gradient length scale, $k_y/\rho_i$ is the Fourier wave number in the direction perpendicular to both the direction of the background gradients $[x\rightarrow k_x/\rho_i]$ and the magnetic field $[z\rightarrow k_z/\rho_i]$, $k_\perp = \sqrt{k_x^2+k_y^2}$, $\bar\phi_{\mathbf k}(\rho_i T_{e0}/L_n e)$ is the gyro-averaged electrostatic potential, $T_{e0}$ is the background electron temperature, $e$ is the elementary charge, and $\nu (v_{ti}/L_n)$ is the collision frequency. We use the Lenard-Bernstein collision operator for the parallel velocity, for which the Hermite polynomials are eigenvectors: $\nu \partial_v [(1/2)\partial_v +v] \rightarrow \nu n $. 

The electrostatic potential is directly proportional to the zeroth-order Hermite polynomial:

\begin{equation}
    \bar \phi_{\mathbf k} = \frac{\pi^{1/4}e^{-k_\perp^2/2} f_{\mathbf k,0}}{1+\tau-\Gamma_0(b)},
    \label{potential}
\end{equation}
%where $\tau$ is the ratio of the ion to electron temperature, and $\Gamma_0(x) = e^{-x}I_0(x)$, with $I_0$ the zeroth-order modified Bessel function.
The first three terms on the right hand side of~\ref{dnaeq} correspond to the gradient drive, the 4$^{th}$ to landau damping, the 5$^{th}$ to phase mixing, the 6$^{th}$ to collisions, and the last is the nonlinearity.
%$\omega_T$ is the normalized inverse temperature gradient scale length, $\omega_n$ is the normalized inverse density gradient scale length, $\nu$ is the collision frequency, and $n$ is the number of the Hermite moment.  The wavenumbers $k_{x,y,z}$ are in the direction of the background gradients, binormal direction, and parallel (to the magnetic field) direction, respectively.  

This system of equations is numerically solved using the DNA code~\cite{Hatch13,jpp2014}.

The phase mixing term, $ik_z[\sqrt{n} f_{\mathbf k, n-1} + \sqrt{n+1} f_{\mathbf k, n+1}]$, depends on $ f_{\mathbf k, n\pm 1}$ and results in the transfer of energy between scales in phase space (see the following section for a detailed discussion).  The dependence of the equation for $ f_{\mathbf k,n}$ on $ f_{\mathbf k, n+1}$ is responsible for the closure problem; the evolution of a given moment depends directly on the next higher order moment, so the set of equations is not closed.  Some truncation strategy is required. The simplest closure scheme is naive truncation: explicitly evolve $n_{\max}$ moment equations, and set $ f_{\mathbf k, n_{\max}+1}=0$. If the system is sufficiently collisional, low-moment truncation is viable~\cite{braginskii_65}.  In a weakly collisional system, if a sufficiently high number of moments is retained, the simulation can be considered to be kinetic and closure by truncation, or via a simple high-n closure~\cite{loureiro_2013}, generally does not disturb the low order moments~\cite{Hatch13,jpp2014}.  If, however, one wishes to evolve a fluid system (i.e. evolve only a few moments), simple truncation will generally produce deviations from the kinetic system, particularly at low collisionality where Landau damping / phase mixing is an important effect.  

%\DRHnote{Akash: please add energy equations here from Hatch PRL 2013 or Hatch JPP 2014.  Make sure notation is consistent with this paper.}
\subsection{Free Energy Equations}
\label{sec:energy}

In order to understand the effects and limitations of various closures, it is useful to conceptualize the turbulent dynamics in the context of an energy equation.  The free energy ~\cite{jpp2014} is given by:
\begin{equation}
\varepsilon_{\mathbf{k}, n}=\varepsilon_{\mathbf{k}}^{(\phi)} \delta_{n, 0}+\varepsilon_{\mathbf{k}, n}^{(f)}
\end{equation}
with field component
\begin{equation}
\varepsilon_{\mathbf{k}}^{(\phi)}=\frac{1}{2}\left(\tau+1-\Gamma_{0}\left(k_{\perp}^{2}\right)\right)^{-1}\left|\phi_{\mathbf{k}}\right|^{2}
\end{equation}
and entropy component
\begin{equation}
\varepsilon_{\mathbf{k}, n}^{(f)}=\frac{1}{2} \pi^{1 / 2}\left|{f}_{\mathbf{k}, n}\right|^{2}.
\end{equation}

The free energy evolution equation can be obtained from \ref{dnaeq} and \ref{potential}:
\begin{equation}
\frac{\partial \varepsilon_{\mathbf{k}, n}^{(\phi)}}{\partial t}=J_{\mathbf{k}}^{(\phi)} \delta_{n, 0}+N_{\mathbf{k}, n}^{(\phi)}
\label{esevolution}
\end{equation}
and
\begin{multline}
\frac{\partial \varepsilon_{\mathbf{k}, n}^{(f)}}{\partial t}=\omega_T Q_{\mathbf{k}} \delta_{n, 2}-C_{\mathbf{k}, n}-J_{\mathbf{k}}^{(\phi)} \delta_{n, 1}\\
+J_{\mathbf{k}, n-1 / 2}-J_{\mathbf{k}, n+1 / 2}+N_{\mathbf{k}, n}^{(f)}.
\label{entropyevolution}
\end{multline}

The terms on the RHS of \ref{esevolution} and \ref{entropyevolution} represent various energy injection, dissipation, and transfer channels. 
The only energy sink---collisional dissipation $C_{\mathrm{k}, n}=2 \nu n \varepsilon_{\mathrm{k}, n}$---is directly proportional to the Hermite number n multiplied by the free energy.
The energy source $ \omega_T Q_{\mathrm{k}}=\omega_T \Re\left[-\frac{\pi^{1 / 4}}{2^{1 / 2}} i k_{y} {f}_{2}^{*} \bar{\phi}\right] $is proportional to the perpendicular heat flux $Q_{\mathbf k}$. 

There are also two conservative energy transfer channels. The nonlinear energy transfer $N_{\mathbf k,n}^{(f)}$ redistributes energy in $k$ space but does not transfer energy between different $n$ and is not a net source or sink (it vanishes under summation in k-space).  

$J_{\mathbf{k}}^{(\phi)}=\Re\left[-i k_{z} \phi^{1 / 4} \bar{\phi}^{*} {f}_{\mathbf{k}, 1}\right]$ is the energy transferred between the field component at $n = 0$ and the entropy component (i.e., Landau damping).

For our purposes of studying closures, the most important terms are the linear phase mixing terms $J_{\mathrm{k}, n-1 / 2}=\Re\left[-\pi^{1 / 2} i k_{z} \sqrt{n} {f}_{\mathrm{k}, n}^{*} {f}_{\mathrm{k}, n-1}\right]$ and $J_{\mathbf{k}, n+1 / 2}=\Re\left[\pi^{1 / 2} i k_{z} \sqrt{n+1} {f}_{\mathbf{k}, n}^{*} {f}_{\mathbf{k}, n+1}\right]$.  These terms also represent a conservative energy transfer channel, albeit in velocity space.  They conservatively transfer energy between $n$ and $n − 1$, $n + 1$ respectively but do not transfer energy in k-space.  One way to characterize the closure problem is determining the proper value of ${f}_{n+1}$ so that $J_{\mathbf{k}, n+1 / 2}$ sends the proper amount of energy to higher order moments---or, as the case may be, \textit{receives} the proper amount of energy \textit{from} higher order moments.  Below, in Sec.~\ref{closure_evaluation}, we will analyze several closures in terms of their capacity to recover the proper (turbulent, kinetic) rates of energy transfer in phase space. 

\section{Closures}
\label{closures}
%Outline:

%- Example set of fluid or gyrofluid equations

%- Brief explanation of motivation behind the HP closure.

%- Write down the HP closure and specify the coefficient values.

%- Specify the relationship between hermite moments in~\ref{reduced_hermite} (relationship between Hermite moments and fluid moments?  More details in appendix?)

%- Mention the Snyder closure and indicate the appendix for a more detailed explanation and derivation of the HP closure

%- Maybe mention the form of the SVD closure so that the plots in the next section with the SVD coefficients make sense 

%\DRHnote{I need to cite some more Hermite papers--Schekochihin, Parker, Watanabe, Hatch, Loureiro}
%\DRHnote{You should at least sketch out the HP and Snyder derivations here even if details are in the appendix.  One or two pargraphs each, perhaps?}

In this section, we describe several closure schemes as applied to our reduced gyrokinetic system.  All closure schemes are of the same class: $ f_{\mathbf k, 4} = \sum_{i=0}^3 A_{\mathbf k, i}  f_{\mathbf k,i}$, i.e., closures that express the last moment in terms of a linear combination of the lower moments. Some closures will have coefficients $A_{\mathbf k, i}$ that are specific to the wave-vector, $\mathbf k$ but others will not, instead having $A_{\mathbf k, i} = A_i$ for all $\mathbf k$.

In a kinetic model where a large number of moments is retained, truncation, which entails setting $ f_{\mathbf k, n_{max}+1} = 0$, can be used.  Alternatively, a simple high-$n$ closure as described in Ref.~\cite{loureiro_2013} can be applied. However, our goal in this work is to formulate a fluid model that captures the relevant kinetic physics while retaining only the most thermodynamically-relevant quantities, namely the first four moments. To achieve this, we require a closure for $ f_{\mathbf k, 4}$ that is more intelligent than simple truncation.  The following subsections describe HP-style closures and the new LMM closure scheme.

\subsection{HP-style closures}
\label{HP_closures}
Here we provide a brief description of HP-style closures.  Derivations and verification of these closures can be found in Appendix A.  The HP closure~\cite{hammet90,Smith97} is designed so that the dispersion relation, also referred to as the kinetic response function, arising from the hierarchy of closed moment equations matches the linear kinetic dispersion relation arising from the Vlasov-Poisson kinetic system. The exact kinetic response function, which involves the plasma dispersion function, $Z(\omega)$, is
\begin{equation}
    R_{00}(\omega) = -iZ(\omega).
    \label{kinetic_response}
\end{equation}
The HP closure for the N$^{th}$ moment takes the form $ f_N = \sum_{i=0}^{N-1} A_i  f_i(\omega)$. Combining this closure ansatz with the hierarchy of moment equations results in an approximate response function $R_{00}^a(\omega)$, a polynomial in $\omega$ involving the closure coefficients, $A_i$.

The HP closure enforces a match in the low frequency limit, $\omega\rightarrow 0$, so the Taylor expansion for plasma dispersion function can be used, which turns Eq.~\ref{kinetic_response} into a polynomial in $\omega$. The closure coefficients $A_i$ can then be chosen so that $R_{00}^a(\omega) = R_{00}(\omega)$. A detailed derivation of the HP closure can be found in Appendix~\ref{hp_appendix}.

The HP closure for the 4$^{th}$ moment in our system is 
\begin{equation}
f_{\mathbf k,4} = sgn(k_z) A_3 f_{\mathbf k,3} + A_2f_{\mathbf k, 2}
\label{HP_closure}
\end{equation}
where the coefficients are $A_3 = -1.759i$ and $A_2 = 0.755$. These coefficients are the same for all $\mathbf k$, so the only $\mathbf k$-dependence for this closure comes from the $sgn(k_z)$.

%The HP closure takes the form $f_{\mathbf k,4} = -i sgn(k_z) A_3\hat f_{\mathbf k,3} + A_2f_{\mathbf k, 2}$ where coefficients $A_3$ and $A_2$ being the same for all $\mathbf k$, so the only $\mathbf k$-dependence coming from the $sgn(k_z)$. In this closure, the coefficients are chosen so that the linear dispersion relation of the closed system matches that of the kinetic system. A detailed derivation of this closure and its coefficients can be found in~\ref{hp_appendix}.

In order to test HP-style closures in collisional regimes, we consider a generalization of the HP closure, developed by Snyder in ~\cite{Snyder97}, which also includes the effects of collisionality. We have developed a collisional extension of the HP closure, the Hammet-Perkins-Collisional (HPC) closure, which is inspired by Snyder's method but modified to match the low frequency limit through second order.

The procedure for arriving at the HPC closure for the N$^{th}$ moment, $ f_{\mathbf k, N}$, is as follows. 
First, use the HP method to determine the closure for the N+1$^{th}$ moment, then substitute this expression for $ f_{\mathbf k, N+1}$ into the linearized time evolution equation for $ f_{\mathbf k, N}$ and take the low frequency limit of this equation ($\partial  f_{\mathbf k,N}/\partial t = 0$). Differentiating this equation with respect to time and then using the low frequency limit of the time evolution equation for the $N-1^{th}$ moment yields a collisional closure for the $N^{th}$ moment.

The HPC closure for the 4$^{th}$ moment in our system is

\begin{multline}
f_{\mathbf k,4} = \frac{-3.051ik_z\nu - 1.759ik_z^2sgn(k_z)}{1.838\nu^2 + 3.709 k_z\nu sgn(k_z) + k_z^2}f_{\mathbf k,3}\\
+ \frac{0.755 k_z^2}{1.838\nu^2 + 3.709 k_z\nu sgn(k_z) + k_z^2}f_{\mathbf k,2}
\label{HPC_closure}
\end{multline}

This is a 2$^{nd}$ order accurate (for small $\omega$) closure for $f_{\mathbf k, 4}$ in terms of $f_{\mathbf k, 3}$ and $f_{\mathbf k,2}$ including collisional effects.
A detailed derivation of this closure and its coefficients can be found in Appendix~\ref{snyder_appendix}.

Note that if one takes the collisionless limit, $\nu\rightarrow0$, of Eq.~\ref{HPC_closure}, the collisionless closure given in Eq.~\ref{HP_closure} is recovered. 

%\begin{equation}
%     f_{\mathbf k,4} = \frac{ -ik_z(2+\sqrt5 A_3) }{ik_zsgn(k_z)\sqrt5 A_4 + 4\nu}  f_{\mathbf k,3}. 
%\end{equation}
%where $A_4 = -1.805i$ and $A_3 = 0.803$. A detailed derivation of this closure and its coefficients can be found in~\ref{snyder_appendix}.

%\DRHnote{need to write out the A's for both HP and Snyder here}
Both the HPC and HP closures were initially designed for models based on the conventional fluid moments in which the $n^{th}$ fluid moment is calculated by integrating the kinetic distribution times velocity to the $n^{th}$ power. Subsequent work generalized the procedure for Hermite-based systems~\cite{Smith97}.  The relationship between the Hermite moments and the fluid moments is very simple and is shown in Appendix~\ref{hermite_appendix}.

%The LMM closures used take advantage of all the lower moments and have $\mathbf k$- specific coefficients: $\hat f_{\mathbf k, 4} = A_{\mathbf k, 3} \hat f_{\mathbf k,3} + A_{\mathbf k, 2} \hat f_{\mathbf k,2} +A_{\mathbf k, 1} \hat f_{\mathbf k,1} +A_{\mathbf k, 0} \hat f_{\mathbf k,0}$. The way these coefficients are calculated will be detailed in Sec.~\ref{svd_closure}.

\subsection{The Learned Multi-Mode Closure}
\label{svd_closure}

%\DRHnote{I need to do more explanation / context in this section}

%As described in the previous section, the HP approach extracts closure coefficients from the kinetic theory of the linear operator.  %In the simplest case, this accounts only for the phase mixing (parallel streaming) term in the linear operator.  More comprehensive treatments may solve for the closure coefficients for the entire linear operator (including collisionality and drive terms).  
We now ask the question of how a closure may be generalized for a nonlinear system in which the turbulent dynamics continually perturb the relationships between the low order moments retained in the system.  

This is motivated, in part, by several recent results showing discrepancies between linear and nonlinear phase mixing dynamics.  Refs.~\cite{plunk2013,kanekar_schekochihin_dorland_loureiro_2015} investigate the effect of a stochastic forcing term on Landau damping rates, demonstrating large deviations from the linear expectations for some parameters.  Refs.~\cite{parker_2016,schekochihin_2016,Meyrand1185} demonstrate a `fluidization' of collisionless plasma turbulence---i.e., a large reduction of Landau damping rates due to the cancellation of the forward velocity space cascade due to turbulence.  Likewise, Ref.~\cite{Hatch_2016} observes Landau damping rates far smaller than the linear predictions in a turbulent system (see Fig. 10 of that paper).  We thus posit that in order to capture the phase mixing rates appropriate for a turbulent, kinetic system, a closure should be endowed with the versatility to adapt to the nonlinear state.  To this end, we propose a closure scheme that learns directly from the turbulent kinetic system. 

To illustrate the closure strategy, consider the Hermite-based system described in Sec.~\ref{reduced_hermite} at two different truncation levels: (1) a four-moment fluid system, and (2) a kinetic system of N Hermite moments, where N is large enough that the system is effectively kinetic (in our simulations we opt for $N=48$).  For a given wavevector, ${\bf k}$, an eigenvector of the linear operator is simply a vector with the complex values of each moment---i.e., a 4D vector in the fluid system and an ND vector in the kinetic system.  

The following closure approach is conceptually similar to the HP approach.  For a given set of physical parameters (gradient drive, collisionality), solve for the linear eigenvector of the kinetic system.  Then use the relationship between $g_4$ and $g_3$ from this kinetic eigenvector to close the fluid system.  If the linear eigenvector persists unmodified in the nonlinear state, this approach would be sufficient.  However, as described above, important nonlinear modifications are observed in turbulent systems.  Consequently, our strategy is to `learn' an appropriate closure directly from the turbulent kinetic system.  

We do so by extracting from a nonlinear kinetic simulation an `optimal' basis for the nonlinear turbulent state at each wave vector.  For a four-field fluid model, we extract this optimal basis from the first five moments of the kinetic system in order to retain the information necessary to close the system.  The turbulent \textit{fluid} state is then projected onto these basis vectors (with the fifth moment of each removed).  Since these basis vectors are attached also to the kinetic information (i.e. the fifth moment), this projection can be used to close the system.  Mathematical details are described in the next subsection.  

Since an optimal basis was extracted from a kinetic simulation, we would expect this procedure to be effective at the parameter point of the kinetic `training' simulation.  The utility of this method, however, will depend on the closure retaining efficacy in some non-negligible parameter domain surrounding the training point.  We demonstrate below that this is the case.  

We call this closure strategy the 'Learned multi-mode' (LMM) closure because (1) it `learns' the closure coefficients from the full turbulent kinetic system, and (2) it employs multiple modes (basis vectors) in order to better capture the dynamical variations in the turbulent state.

We end this section by noting some connections with other lines of research.  First, this closure approach is related to various strategies for projection-based model reduction~\cite{Sirovich_87, Berkooz_93, Rozza_2008, Feldmann_95, freund_2003, Peherstorfer_2016}, wherein basis vectors are extracted (often via SVD) from data describing a complex system to reduce the complexity of the underlying models.

We also note some connections with the closure proposed in Refs.~\cite{sugama2001,sugama2003}.  This closure scheme employs two modes (the ITG mode and its complex conjugate) in order to enforce a  `non-dissipative' closure---i.e., it eliminates any energy transfer between the fluid moments and higher order moments.  Consequently, it produces damping rates that are far below  (i.e., zero) the linear values, qualitatively similar to the nonlinear results cited above.  However, the true turbulent system allows energy to shift dynamically between lower and higher order moments.  Consequently, we view this closure as a compelling idea, but one that is perhaps too restrictive. 

We also note the connection between the LMM closure and the line of research exploring the role of damped eigenmodes in plasma microturbulence~\cite{terry_2006,PhysRevLett.106.115003,Hatch2011,Hatch_2016,whelan_2018}, which shows that multiple modes co-existing at a single wavevector play a crucial role in turbulent energetics.  Our 'multi-mode' closure also acknowledges the activity of multiple eigenmodes per wavevector and defines the closure coefficients in terms of the relative amplitude of these modes in the nonlinear state.

\subsection{Implementation of the LMM Closure}
\label{implementation}
%Motivated by the results in the previous section, here we seek a flexible closure for phase mixing in a turbulent system.  More specifically, we seek to accurately resolve the low-order moments, $\hat f_{0:3}$ that define the physical quantities of interest and determine transport fluxes, without retaining the higher order moments $\hat f_{4:\infty}$. 
%\DRHnote{I kind of think we should drop the hats on all the f's.  A bit too much clutter.  What do you think?}
Here we describe the mathematical details of the approach outlined in the previous section.  The closure requires a nonlinear kinetic simulation to formulate a set of basis vectors.  In our case, we use 48 Hermite moments for the full kinetic simulation.  Any number of subsequent fast fluid simulations can then be run requiring explicit computation of only $ f_0$, $ f_1$, $ f_2$, and $ f_3$. In this section we will use bold uppercase letters to denote matrices and bold lowercase letters to refer to vectors.

The full kinetic simulation is used as follows.  Let $\mathbf F_{N\times M}$ ($M$ is the number of time points and $N$ is number of moments retained in the fluid model plus one) be the matrix created from the simulated distribution function at a single wave vector.  The distribution function at a single wavevector is written $ f_i(t)$, where $i = 0,1,...,N-1$ denotes the Hermite number and $t$ takes on discrete values $t_j$ with $j = 0,1,...,M-1$ (the wave vector is suppressed for clarity), so that element $ij$ of $\mathbf F$ is $\mathbf F_{ij}= f_{i}(t_j)$:
%\begin{equation}
%    G =
%    \begin{bmatrix}
%        g_{0}(t_0) & g_{0}(t_1) & \dotsb & g_{0}(t_M) \\
%        g_{1}(t_0) & g_{1}(t_1) & \dotsb & g_{1}(t_M) \\
%        g_{2}(t_0) & g_{2}(t_1) & \dotsb & g_{2}(t_M) \\
%        g_{3}(t_0) & g_{3}(t_1) & \dotsb & g_{3}(t_M) \\
%        g_{4}(t_0) & g_{4}(t_1) & \dotsb & g_{4}(t_M) \\
%    \end{bmatrix}
%    \label{data matrix}
%\end{equation}
\begin{equation}
    \mathbf F =
    \begin{bmatrix}
         f_{0}(t_0) &  f_{0}(t_1) & \dotsb &  f_{0}(t_{M-1}) \\
         f_{1}(t_0) &  f_{1}(t_1) & \dotsb &  f_{1}(t_{M-1}) \\
        \vdots & \vdots & \ddots& \vdots \\
         f_{N-1}(t_0) &  f_{N-1}(t_1) & \dotsb &  f_{N-1}(t_{M-1}) \\
    \end{bmatrix}
    \label{data matrix}
\end{equation}
The SVD of $\mathbf F$ is given by 
\begin{equation}
\mathbf F_{N\times M}=\mathbf U_{N\times N} \mathbf \Sigma_{N\times N} \mathbf V^H_{N\times M}
\label{svd}
\end{equation}
where $\mathbf U$ and $\mathbf V$ are unitary and $\mathbf \Sigma$ is diagonal with real entries. General background information about this extremely useful matrix decomposition and be found in Ref.~\cite{golub2013matrix} and a review on its application to turbulence as proper orthogonal decomposition (POD) can be found in Ref.~\cite{Berkooz}. The columns of the matrix $\mathbf U$ are called the left singular vectors.  In our application, they define $N$ basis vectors for the distribution function.  The rows of $\mathbf V^H$ are the time traces of the amplitude of each of these vectors.  The diagonal entries in $\mathbf \Sigma$ define the singular values, which encompass all the amplitude information.  The utility of the SVD lies in its property that the outer product between the first basis vector and the first time trace (weighted by the corresponding singular value) reproduces more of the fluctuation data (as measured by the Frobenius norm) than any other possible decomposition of this form.  Likewise the superposition of the first two ($n$) outer products captures more of the fluctuation data than any other rank two ($n$) decomposition and so forth. For convenience, we define a matrix $\mathbf{B}$, which weights the basis vectors by their corresponding singular values so that they include the amplitude information: $\mathbf B=\mathbf U\mathbf \Sigma$.
%, and let $\{\mathbf a_1,\mathbf a_2,..., \mathbf a_N\}$ and $\{\mathbf b_1,\mathbf b_2,...\mathbf b_N\}$ be the rows and columns of $B$ respectively.

For the purposes of our desired four moment model, we select $N=5$ (i.e. only a small subset of the 48 total Hermite moments).  Since different Hermite moments are only connected to their direct neighbors, this is sufficient to fully exploit the information in the simulation defining the natural (kinetic, turbulent) relations between $ f_3$ and $ f_4$.

Let $\mathbf f$ represent the column vector of the first four moments at a single time step: $\mathbf f = [ f_0\;  f_1\;  f_2\;  f_3]^T$. In each time step of a subsequent fluid simulation, we numerically advance $\mathbf f$ explicitly via Eq.~\ref{dnaeq}.  The truncated moment, $ f_4$, is calculated as follows.  First, we project the state vector $\mathbf f$ onto the basis formed by the columns of $\mathbf B$. This entails finding the projection coefficients that define the amount of each SVD mode in the turbulent state at a given point in time.  We will call the column vector containing these projection coefficients $\mathbf c$.  We can do this by removing the row corresponding to the unknown $N^{th}$ moment (the 5$^{th}$ row) from $\mathbf B$  and extracting $\mathbf c$ from the following equation:
\begin{equation}
    %\hat f_{0:3} = B_{0:3,0:4}\mathbf c
    \mathbf f = \mathbf M \mathbf c
    \label{svd0}
\end{equation}
where $\mathbf M$ denotes the submatrix of $\mathbf B$ consisting of the first 4 rows and all 5 columns of $\mathbf B$, i.e., $\mathbf M$ is the submatrix produced by removing the last (5$^{th}$) row of $\mathbf B$.
This gives
\begin{equation}
%c = (B_{0:3,0:4})^\dag \hat f_{0:3}
\mathbf c = \mathbf M^\dag \mathbf f
    \label{svd1}
\end{equation}
where $^\dag$ denotes the pseudo-inverse.

Now that we have $\mathbf c$, a length $N$ vector of the inferred mode amplitudes, we can predict $ f_{4}$ by applying these mode strengths to the previously removed row of $\mathbf B$, $\mathbf b_5$. This gives
\begin{equation}
 f_{4} = \mathbf b_5 \mathbf c = \mathbf b_5\mathbf M^\dag \mathbf f = \mathbf c_{LMM}\mathbf f
    \label{svd closure}
\end{equation}
where $\mathbf b_5$ is the 5$^{th}$ row of $\mathbf B$ and $\mathbf c_{LMM} = \mathbf b_5 M^\dag$ is the vector containing the 4 LMM closure coefficients. This procedure is repeated at each wavevector $\mathbf k$ to obtain a full set of coefficients that can be used to conduct an LMM-closed simulation. % - $\mathbf c_{LMM}$ is different for each wavevector.

This procedure results in a closure that reflects the natural relations between moments in the turbulent kinetic system and adapts to the relative amplitude of each basis vector in the nonlinear state.

Regarding computational cost, the LMM closure comes down to the dot product between two length 4 vectors: the closure coefficients, $c_{LMM}$, and the lower order moments, $\mathbf f$. The closure coefficients are computed ahead of time and saved to a file, which is loaded at the beginning of the simulation. During the simulation, the computational expense of the LMM closure is very similar to that of the HP closure; the HP closure requires two complex multiplications per wavevector per time step (one for each of the two HP closure coefficients), and the LMM closure requires four complex multiplications per wavevector per time step.  This is much less demanding than the pseudo-spectral computation of the nonlinearity, so the increased expense is negligible.  The main additional expense is in running nonlinear kinetic simulations for training.  If this can be done sparsely, then the LMM closure is viable.

%Although some extra computational expense is required to take the SVD of the kinetic simulation data and obtain the closure coefficients, this is on the order of the other terms in the linear operator and much less demanding than the pseudo-spectral computation of the nonlinearity. 

%In fact, the projection is only slightly more expensive than the HP closure.  The HP closure requires 2 complex multiplications ($A\cdot \hat f_3$ and $B\cdot \hat f_2$)  per wave vector ($k$) per time step. The LMM closure amounts to a dot product between two length 4 vectors because in Eq. \ref{svd closure}, $B_{4,0:4} (B_{0:3,0:4})^\dag$ is a 1x4 vector times a 5 by 4 matrix which results in a 1x4 vector.  This product is computed ahead of time and saved to a file which is loaded at the beginning of the simulation. In the simulation, this 1x4 vector must be dotted with $\hat f_{0:3}$, which is a 4x1 vector, to get the closure for $\hat f_4$. Thus, the LMM closure requires 4 complex multiplications per wave vector per time step.

\section{Preliminary Closure Tests}
\label{closure_evaluation}
In this section we probe the properties of several closures in comparison with the kinetic system in both linear and nonlinear scenarios.

\begin{figure}[H]
    \centering
    \includegraphics[width=\columnwidth]{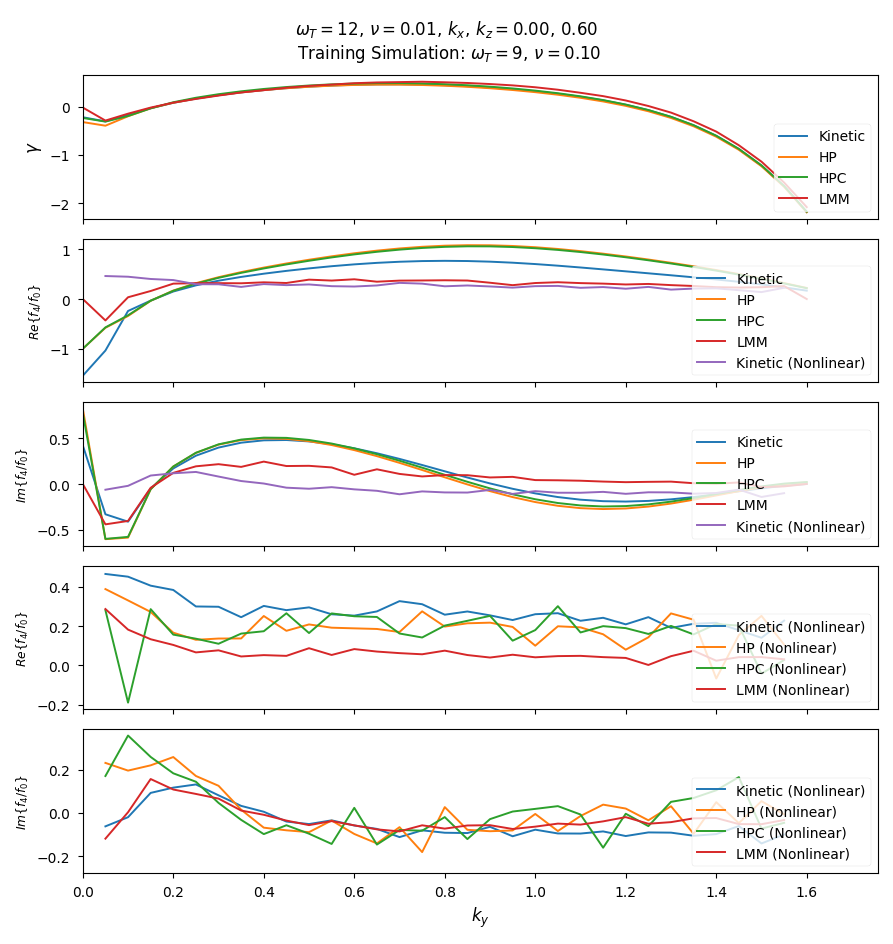}
    \caption{
    \label{linear_agreement}
    Linear growth rates (top panel) and real (2$^{nd}$ panel) and imaginary (3$^{rd}$ panel) parts of $ f_{4}$ normalized to $ f_0$ produced by solving the eigenvalue problem given by the linearized version of Eq.~\ref{dnaeq} plotted against $k_y$ for temperature gradient drive ($\omega_T$) = 12, collision frequency ($\nu$) = 0.01 and $k_x,\,  k_z\, = 0,\, 0.6$. The eigenvalue problem is solved using the linear 48-moment (kinetic) system and also using the 4 moment system closed with the HP closure, HPC closure, and LMM closure.
    %In the kinetic case, $ f_4$ is the 5$^{th}$ element of the eigenvector associated with growth rates plotted in the top panel. In the closed cases, $ f_4$ is calculated as $ f_4 = \sum_{i=0}^3 A_i f_i$ where $A_i$ depend on which closure is being used and $ f_i$ are the lower elements of the eigenvector from the kinetic case. 
    The LMM closure coefficients used to produce this figure were extracted from the kinetic simulation at parameter point ($\omega_T,\nu = 9, 0.1$). Panels 2 and 3 also show the time averaged value of $ f_4 /  f_0$ from the nonlinear kinetic simulation.
    Panels 4 and 5 show the real and imaginary parts of the time averaged value of $f_4/f_0$ from nonlinear kinetic simulations as well as nonlinear LMM,HP, and HPC simulations.
    %\ASnote{Please Check that this caption is clear}
    }
\end{figure}

The HP closure has been shown to faithfully reproduce kinetic Landau damping rates and linear growth rates.  We reproduce this result for our system: simulations exhibit good agreement between kinetic linear growth rates and fluid growth rates using the HP closure.  A representative example is shown in Fig.~\ref{linear_agreement} (top panel), where it is seen that the HP closure, the HPC closure, and the LMM closure all reproduce the growth rates of the linear kinetic system. Growth rates are produced by solving the linearized eigenvalue problem given by Eq.~\ref{dnaeq} for the 48-moment (kinetic system) and the 4-moment fluid system with each of the closures.

Fig.~\ref{linear_agreement} (lower panels) also shows a simple test of the eigenmode structures by plotting the 4$^{th}$ moment, $f_4$, normalized to the zeroth moment. 
These plots are highly relevant since ratios of moments are closely connected to the closure problem.  
In the 2$^{nd}$ and 3$^{rd}$ panels, we plot the real and imaginary parts of this quantity for the 48-moment (kinetic) linear system, the 48-moment nonlinear system, as well as all of the linear 4-moment systems closed by the HP, HPC, and LMM closures. 
In the 4$^{th}$ and 5$^{th}$ panels, we plot the real and imaginary parts of this quantity for the the 48-moment nonlinear system, as well as all of the nonlinear 4-moment systems closed by the HP, HPC, and LMM closures.
For this example, the LMM closure was trained at parameter point $\omega_T = 9, \nu = 0.1$. 
%\DRHnote{So it was trained and applied at the same parameter point?}\ASnote{Yes that's right.}

While the growth rates produced by all the closed systems match the linear kinetic growth rates very closely, the agreement is not as good in the plots of these moment ratios in the 2$^{nd}$ and 3$^{rd}$ panels. $ f_4$ produced by the HP and HPC closures, which are both based on the linear system, exhibit a similar shape to the linear kinetic $ f_4$. However, $ f_4$ of the nonlinear kinetic system exhibits a significantly different shape in $k_y$: the ratio is much smaller.  This is closely mirrored by $ f_4$ produced by the linear system closed by the LMM closure. The capacity of the LMM closure to reproduce the nonlinear result is perhaps unsurprising, as it is based on the nonlinear system.

The 4$^{th}$ and 5$^{th}$ panels show that the ratios of $f_4$ to $f_0$ in nonlinear HP and HPC simulations matches the nonlinear kinetic simulation much more closely than the ratios from the linear HP and HPC systems. Apparently, the HP approach retains the capacity to adapt to the nonlinear state, which will be discussed further below.  The imaginary part of the ratio from the nonlinear LMM-closed system matches the kinetic ratio the best of the three closures shown, but the real part of this ratio is consistently smaller than the kinetic ratio for all $k_y$ shown.

This is an initial indication that dynamics of the linear and nonlinear systems are quite different, consistent with the literature discussed above~\cite{plunk2013,kanekar_schekochihin_dorland_loureiro_2015,Hatch_2016, Meyrand1185}. 

%This motivates the use of a closure like the LMM closure directly targeted at capturing nonlinear effects.  
%\DRHnote{define the wavenumbers for this figure, show only positive wavenumbers, and limit plot range to showing mostly unstable modes}.

%\begin{figure}[H]
%    \centering
%    \includegraphics[width=\columnwidth]{gamma_omt12_nu0.01.png}
%    \caption{
%    \label{linear_agreement}
%    Linear growth rates ($\gamma$) produced by eigenvalue calculations using the fluid model with the HP closure and the fully kinetic model at the most unstable wavenumber, $k_x,\,  k_z\, = 0,\, 0.6$, plotted against $k_y$, for temperature gradient drive ($\omega_T$) = 12, collision frequency ($\nu$) = 0.01.
%    }
%\end{figure}

\section{Nonlinear Closure Tests}
\label{simulation_results}

In order to more thoroughly examine closure performance, simulations covering a wide range of temperature gradients, $\omega_T$, and collision frequencies, $\nu$, were conducted with a fully (reduced gyro-) kinetic model (48 moments : $n_{max}=48$), simply truncated model (4 moments, 5$^{th}$ is set to 0), the standard Hammett-Perkins (HP) closure retaining 4 moments, the HPC closure, and three different LMM closures. 

The scan covers $\omega_T = 5,6,7,8,9,12,15$, and $\nu = 0.01,0.05,0.1,0.2,0.5$. All closures perform very poorly in the completely collisionless regime, $\nu =0$.  This is currently under investigation and may require a more careful treatment of the dissipation in the kinetic system, which is left for future work (collisionless results are shown in Appendix~\ref{flux_appendix}).  Other simulation parameters used are $\omega_n = 1$ and $\tau = 1$.    
%\DRHnote{I would recommend plotting the nu=0 cases separately in the appendix--i.e. not attached to the other simulation restuls}
%\ASnote{Ok, did this for the time traces. I think it makes most sense to keep the error plots as a full grid though because you can see how the nu = 0 errors are so much larger than the others.}
%Versions of these figures including $\nu=0$ can be found in Appendix~\ref{flux_appendix}. 

%~\ASnote{Dr. Hatch add hyperviscosity and hypercollisionality and other terms}.
The grids used in $k-$space are $k_{x,min} = 0.05$, $k_{x,max} = 1.5$, $k_{y,min} = 0.05$, $k_{y,max} = 1.5$, and $k_{z,min} = 0.1$, $k_{z,max}=3.6$.  Hyper-collisions of the form $\nu_h (n/n_{(max)})^8f_n$ are included in the full kinetic simulations in order to enforce decaying moments at high $n$.  Hyper-diffusion of the form $\nu_\perp (k_{x,y}/k_{x,y}^{(max)})^8$ is included as a small scale dissipation mechanism (intended to roughly account for, e.g., nonlinear perpendicular phase mixing). We use $\nu_h = 0.1$ and $\nu_\perp$ = 1 in our simulations.  In addition, a Krook term is applied to the zero and minimum finite $k_z$ modes in order to avoid slow growth of these modes that fails to saturate.  Refs.~\cite{Hatch13,jpp2014} describe such numerical considerations in more detail. 
%\DRHnote{add the specific values of the coefficients that you used in simulations}

For reference, the exact HP closure used was $ f_{\mathbf k,4} = 0.755  f_{\mathbf k, 2} -i (1.759 sgn(k_z))  f_{\mathbf k, 3}$.
The HPC closure used was
%\begin{equation}
%    f_{\mathbf k,4} = \frac{ -ik_z(2+\sqrt5(.803))}{ik_zsgn(k_z)\sqrt5 (-1.805i) + 4\nu} f_{\mathbf k,3}. 
%\end{equation}
\begin{multline}
f_{\mathbf k,4} = \frac{-3.051ik_z\nu - 1.759ik_z^2sgn(k_z)}{1.838\nu^2 + 3.709 k_z\nu sgn(k_z) + k_z^2}f_{\mathbf k,3}\\
+ \frac{0.755 k_z^2}{1.838\nu^2 + 3.709 k_z\nu sgn(k_z) + k_z^2}f_{\mathbf k,2}
\end{multline}
The derivation of these coefficients is described in Appendices ~\ref{hp_appendix} and ~\ref{snyder_appendix}.

A key question for the LMM closure is the parameter domain over which the closure remains viable.  We would expect the applicability of a set of LMM closure coefficients to deteriorate as the distance in parameter space from the training simulation increases. %Thus, when a large parameter grid is used, it is necessary to perform more than one kinetic training simulation in order to have LMM closures that are accurate everywhere on the grid. 
Of course, the computational expense of kinetic simulations requires that the number of training simulations be kept minimal in order for the closure to be useful. In order to probe the question of how far the closure applies throughout parameter space, we selected 3 kinetic training simulations spread throughout the parameter grid. %The results presented in this section illustrate how LMM closure performs at different parameter points at various distances from the training point. %Constructing several LMM closures as we have done, and using the appropriate LMM closure at each parameter point (the closure trained closest to that point), results in great performance.

The three different LMM closures are obtained by applying the method described in Sec.~\ref{implementation} to extract coefficients from kinetic simulations at $\omega_T, \nu = 6,0.01$, $\omega_T, \nu = 9,0.1$, at $\omega_T, \nu = 12,0.5$.  We refer to these three LMM closures as LMM-Left, LMM-Middle, and LMM-Right respectively, indicating the region of the scanned parameter space within which their training simulation lies. %`Training simulation' refers to the kinetic simulation from which the LMM closure coefficients were extracted. 

\subsection{Tests of Energy Dissipation}

In order to gain insight into the nonlinear dynamics and their effect on the closure problem, we investigate the energy evolution equation, Eq.~\ref{entropyevolution}. In Eq.~\ref{entropyevolution}, the contribution from phase mixing defines the energy flux to higher order moments~\cite{jpp2014}.
More specifically, $\chi_{n+1/2} \equiv J_{n+1/2}/(|k_z||f_n|^2)$, the normalized rate at which energy is transferred to/from higher order moments (the phase mixing rate), is defined by a correlation between two neighboring moments.
The linear physics defines a fixed, dissipative, relationship between $ f_n$ and $ f_{n+1}$.  In the presence of turbulence, however, the various moments are continually perturbed by the nonlinearity, resulting in correlations that can differ substantially from the linear expectation.  

\begin{figure}[H]
\captionsetup[subfigure]{justification=centering}
\centering
\begin{subfigure}{0.5\textwidth}
\includegraphics[width=\textwidth]{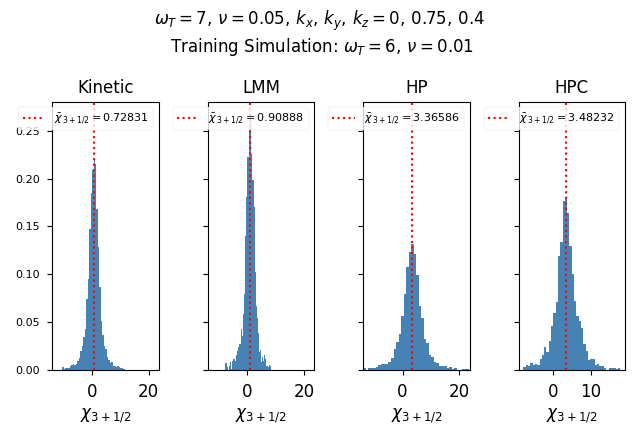}
\caption{}
\end{subfigure}
\begin{subfigure}{0.5\textwidth}
\includegraphics[width=\textwidth]{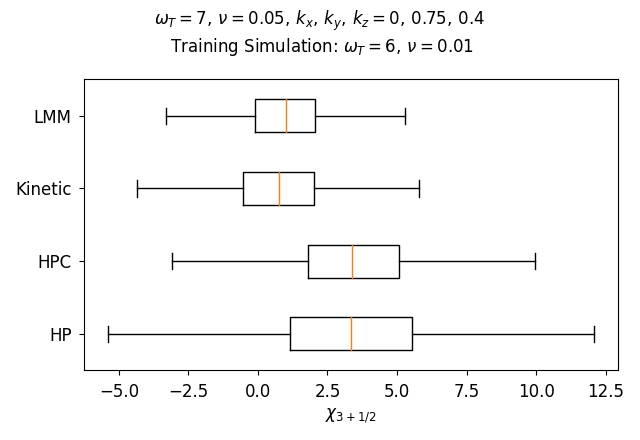}
\caption{ }
\end{subfigure}
\caption{
\label{boxplot}
 Probability distribution functions (a) and box and whisker plots (b) showing the distribution of $\chi_{3+1/2}$, the energy transferred between the 3$^{rd}$ and 4$^{th}$ moments, in the Kinetic, LMM-closed, HPC-closed, and HP-closed systems for  $\omega_T = 7$ and $\nu = 0.05$ at the most unstable wavevector, $k_x = 0$, $k_y = 0.75$, $k_z = 0.4$. The LMM coefficients used to produce this plot were extracted from the kinetic simulation at $\omega_T, \nu = 6, 0.01$. Red dashed lines show the average value of each distribution, $\bar \chi_{3+1/2}$. 
 %Positive values of $J_{3+1/2}$, energy transferred from moment 3 \textit{to} moment 4, indicate dissipation. It can be seen clearly here that the HP closure results in much more dissipation than is present in the kinetic system, while the LMM closure much more accurately captures these energy dynamics. The average value of $J_{3+1/2}$ from the HP closure is much too large, while the average value from the LMM closure almost exactly matches the average from the Kinetic case. Additionally, the shape of the distribution for the LMM-closed case is very similar to that of the kinetic distribution while the HP distribution is heavily skewed to the right.
}
\end{figure}

These considerations are illustrated in Fig.~\ref{boxplot}, which shows the distribution (accumulated over time) of the energy transfer rate between the 3$^{rd}$ and 4$^{th}$ moments, $\chi_{3+1/2}$, for kinetic, LMM-closed, HP-closed, and HPC-closed simulations. The average dissipation, $\bar \chi_{3+1/2}$ resulting from the HP and HPC closures is much larger than the dissipation present in the kinetic system.
We note that the HP and HPC closures would likely perform better by this metric with the inclusion of more moments, which may be explored in future work.  The LMM-closure, being based on the nonlinear system, produces dissipation that matches the kinetic level quite closely. We note that the LMM closure coefficients used are extracted from a training simulation with different parameters than the simulation being examined (both parameter points are noted in the title of the figure), which indicates the effectiveness of the LMM closure even in simulations with parameters different from the training parameters.

\begin{figure}[H]
\centering
\includegraphics[width=0.5\textwidth]{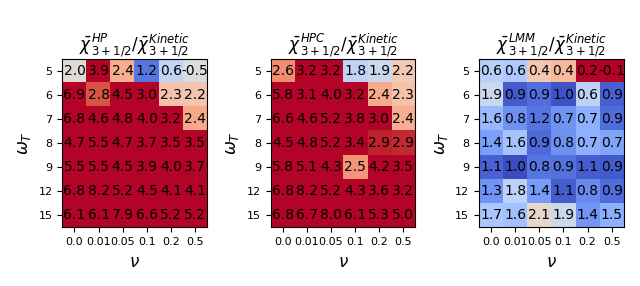}
\caption{
\label{jtile}
Ratios of the average value of $\chi_{3+1/2}^{closed}/\chi_{3+1/2}^{Kinetic}$  at the most unstable wavevector throughout parameter space for the HP, HPC, and LMM closures. Values below 1 indicate not enough dissipation and values above 1 indicate too much dissipation. 
This heatmap is set up such that fractions far from 1 in either direction are penalized the same way. For example, ratios of 0.5 and 2 will be the same color.
As shown here, the LMM closure matches kinetic dissipation levels much better than the HP and HPC closures throughout most of our parameter grid.
%As shown here, the HP and HPC closures perform better in the top left region of the grid and the LMM closure matches kinetic dissipation levels better throughout the rest of the parameter grid.}
%Three different sets of LMM closure coefficients are used to produce this plot: one set is extracted from the kinetic simulation at $\omega_T, \nu = 6, 0.01$, one from $\omega_T, \nu = 9,0.1$, and one from $\omega_T, \nu = 12,0.5$. These three different LMM closures are detailed in sec.~\ref{simulation_results} and are referred to as the LMM-Left, LMM-Middle, and LMM-Right closures respectively. They are named as such to indicate the region of our parameter space in which they were trained. At each point in our parameter grid, we use the LMM closure that was trained closest to that grid point to produce the values of $\bar J^{LMM}_{3+1/2}$ shown in this plot. Table~\ref{opt_table} in Appendix~\ref{flux_appendix} shows which of these closures (LMM-Left,Right, or Middle) was used at each grid point.
%\ASnote{Dr. Hatch please check this caption for clarity. Especially about which closure was used at each grid point.}
}
\end{figure}

%\DRHnote{need to state the training points for each of the LMM cases} 
Fig.~\ref{jtile} shows the ratio of the average value of $\chi^{Closed}_{3+1/2}$ to the average value of $\chi^{Kinetic}_{3+1/2}$ at the most unstable wavevector for the HP, HPC, and LMM closures for every point in the extended parameter scan.
%Three different LMM closures (LMM closures with three different training points) are used to produce the  plot on the right in Fig.~\ref{jtile}. One was trained on the kinetic simulation at $\omega_T, \nu = 6, 0.01$, one on the kinetic simulation at $\omega_T, \nu = 9,0.1$, and one on the kinetic simulation at $\omega_T, \nu = 12,0.5$. These three different LMM closures are detailed in sec.~\ref{simulation_results} and are referred to as the LMM-Left, LMM-Middle, and LMM-Right closures respectively. They are named as such to indicate the region of our parameter space in which they were trained. 
Three different sets of LMM closure coefficients are used to produce this plot: one set is extracted from the kinetic simulation at $\omega_T, \nu = 6, 0.01$, one from $\omega_T, \nu = 9,0.1$, and one from $\omega_T, \nu = 12,0.5$. These three different LMM closures are detailed in sec.~\ref{simulation_results}.  %and are referred to as the LMM-Left, LMM-Middle, and LMM-Right closures respectively. They are named as such to indicate the region of our parameter space in which they were trained. 
At each point in our parameter grid, we use the LMM closure that was trained closest to that grid point to produce the values of $\bar \chi^{LMM}_{3+1/2}$ shown in this plot.
%Table~\ref{opt_table} in Appendix~\ref{flux_appendix} shows which of these closures (LMM-Left,Right, or Middle) was used at each grid point.
%At each point in our parameter grid, we use the LMM closure that was trained closest to that grid point to produce the values of $\bar J^{LMM}_{3+1/2}$ shown in this plot. Table~\ref{opt_table} in Appendix~\ref{flux_appendix} shows which of these closures (LMM-Left,Right, or Middle) was used at each grid point.

The HP and HPC closures overestimate dissipation levels throughout most of the parameter grid.
%, but do fairly well in the top left of the gird at low collisionality and gradient drive. 
They perform best at low gradient drive ($\omega_T$) with deteriorating performance as gradient drive is increased. %Interestingly, the LMM closure shows the opposite behavior for this test; it severely underestimates dissipation levels at low collisonality and gradient drive where HP and HPC do well, but performs well elsewhere.}
The LMM closure matches kinetic dissipation levels significantly better than both HP closures throughout the parameter space.

%\textcolor{orange}{ All of the simulations for which the LMM closure does poorly use LMM coefficients extracted from the $\omega_T=6,\nu=0.01$ simulation as shown in Table~\ref{opt_table}. It appears that the coefficients extracted from this simulation are not dissipative enough. The other points on the grid use LMM coefficients extracted from either the $\omega_T=9, \nu=0.1$ or $\omega_T=12,nu=0.5$ kinetic simulations and quite accurately reproduce kinetic dissipation levels. }

It is important to note that the substantial disagreement between kinetic and HP/HPC dissipation levels is somewhat misleading. The discrepancy between the heat flux saturation levels examined in the next section is much less than is suggested by the large difference in phase mixing rates. 

\begin{figure*}[ht]
    \centering
    \includegraphics[width=\textwidth]{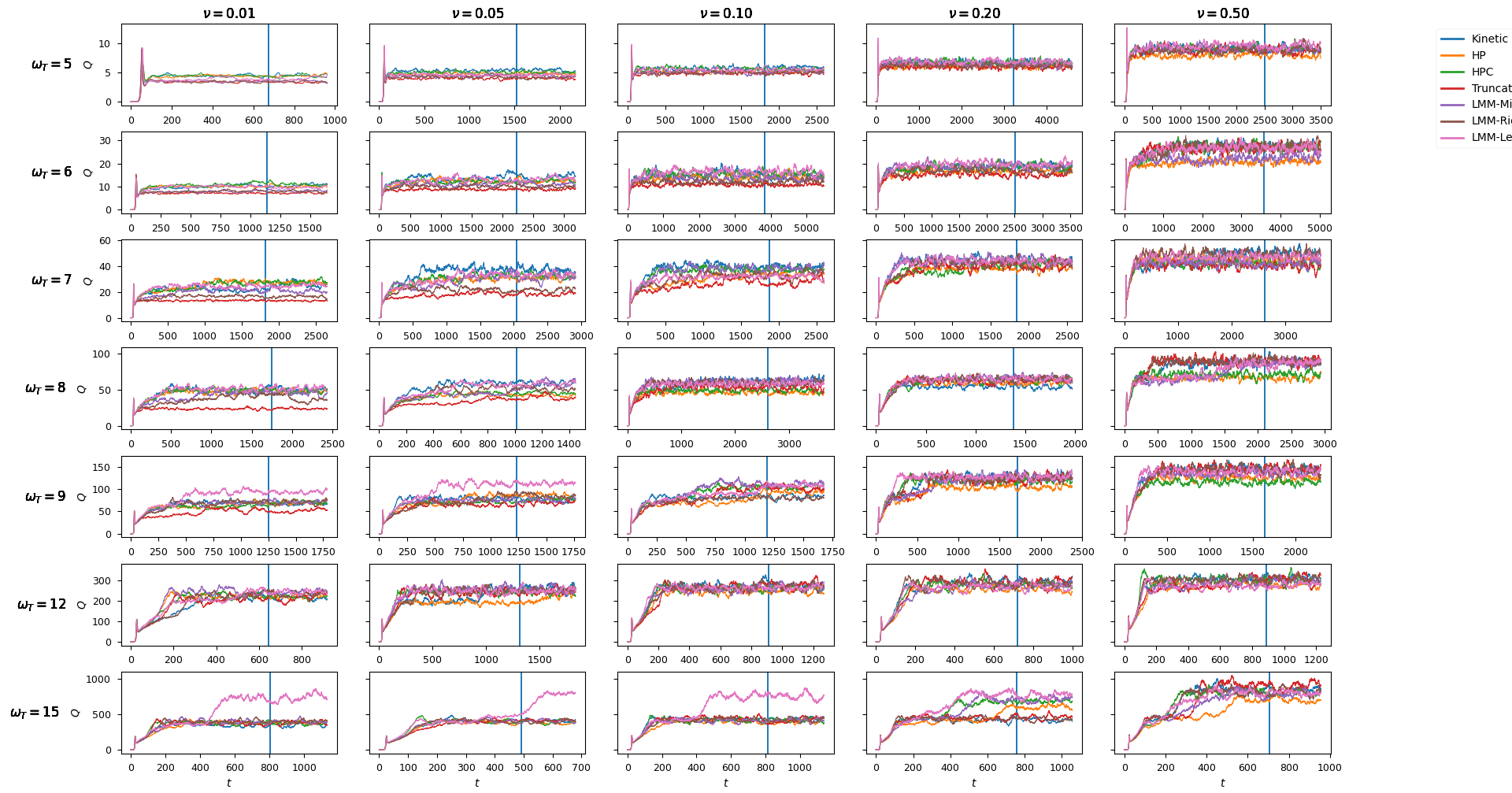}
    \caption{
        \label{flux}
        Time traces of the total radial heat flux ($Q$) for Kinetic (blue), HP (orange), HPC (green), truncated (red), LMM-Middle (purple), LMM-Right (brown), and LMM-Left (pink)  simulations for temperature gradient drives ($\omega_T$) ranging from 5 to 15 (increasing downward by panel) and collision frequencies ($\nu$) ranging from 0.01 to 0.5 (increasing to the right by panel). The metric of performance is the final saturation level. The vertical blue lines show the cutoff point - 70\% of the simulation time - after which each heat flux curve is averaged to get the final saturation level. 
    }
\end{figure*}

\subsection{Comparison of Heat Fluxes}

Ultimately, we would like closed simulations to reproduce the most important macroscopic behavior of gyrokinetic simulations, notably the saturated value of the radial heat flux, $Q = \sum_{k_x,k_y,k_z} Q_{\mathbf k}$, where $Q_{\mathbf k}$ is defined in the discussion surrounding Eq.~\ref{entropyevolution}.  We view this metric---the proximity of the saturated heat flux for a given closure scheme to that of the kinetic simulation---to be the most relevant metric for the performance of the closure.

\begin{figure*}[ht]
    \centering
    \includegraphics[width=\textwidth]{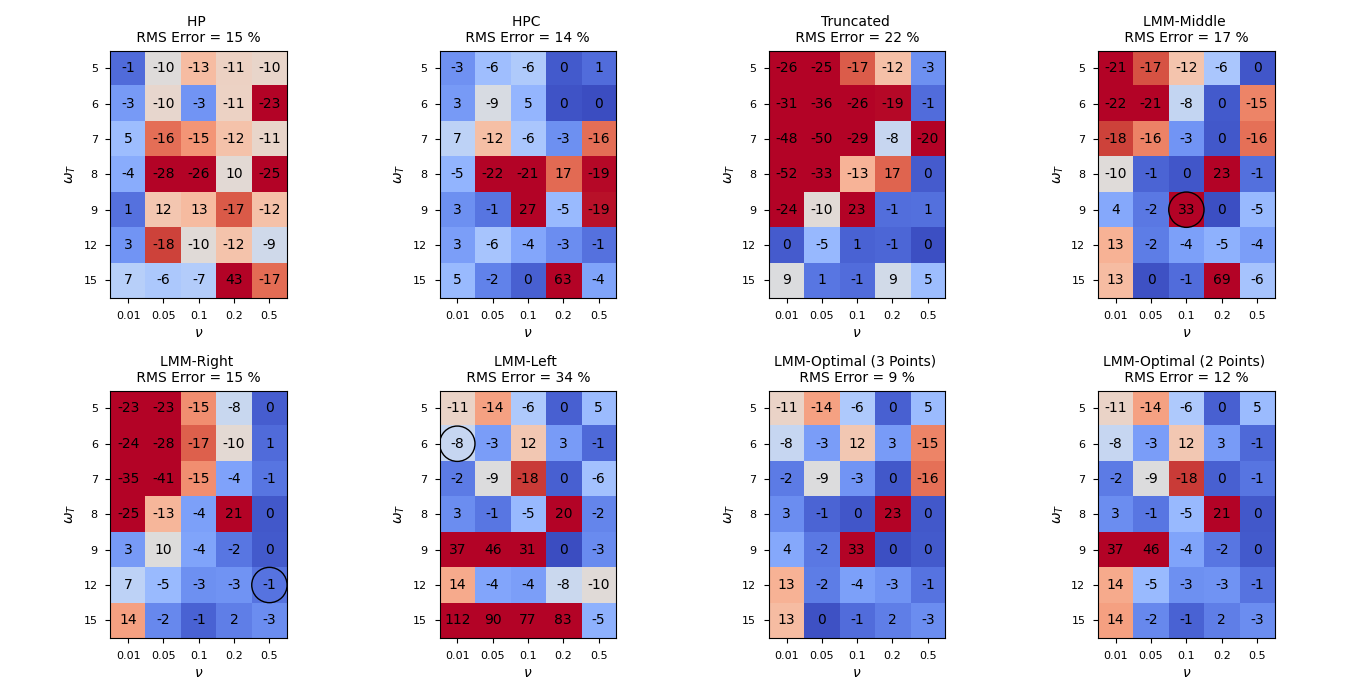}
    \caption{
        \label{error}
        Percent error in saturated heat flux for each closure (HP, HPC, Truncation, LMM-Middle, LMM-Right, LMM-Left, LMM-Optimal) as compared against the kinetic simulation throughout parameter space. The circles on the 4$^{th}$, 5$^{th}$, and 6$^{th}$ figures indicate which simulations from which  the LMM coefficients were extracted. Percent errors are calculated as $(Q^{Closed} - Q^{Kinetic})/Q^{Kinetic} \times 100$ where $Q^{closed}$ is calculated by averaging the last 30\% of the time trace of the heat flux from the closed simulation and $Q^{Kinetic}$ is calculated by averaging the last 30\% of the time trace of the heat flux from the kinetic simulation. 
        %The average error in for each closure is calculated by averaging the absolute value of the errors.
        The RMS error for each closure is also shown above each plot.
        The 7$^{th}$ plot, LMM-Optimal (3 Points), displays error of the LMM closure trained at the nearest parameter point using three training points. The 8$^{th}$ plot, LMM-Optimal (2 Points), displays the error of the LMM closure trained at the nearest parameter point using only the left and right training points.
        %For this case, at each grid point, the LMM closure trained at the parameter point closest to that grid point is used. For example, at $\omega_T, \nu = 5, 0.05$,  the LMM-Left closure with coefficients extracted from the $\omega_T, \nu = 6,0.01$ kinetic simulation is used and at $\omega_T, \nu = 15,0.5$, the LMM-Right closure with coefficients extracted from the $\omega_T, \nu = 12,0.5$ kinetic simulation is used. Table~\ref{opt_table} in Appendix~\ref{flux_appendix} shows which training simulation was used to run the LMM-closed simulation at each grid point for the LMM-Optimal plot.
        %At each grid point, the LMM closure trained at the parameter point closest to that grid point is used. Table~\ref{opt_table} in Appendix~\ref{flux_appendix} shows which training simulation was used to run the LMM-closed simulation at each grid point.
    }
    \end{figure*}

Time traces of the heat fluxes produced in the kinetic simulation and seven closed simulations are shown for each combination of input parameters, $\omega_T$ and $\nu$, in Fig.~\ref{flux}.  
The final saturation levels of each simulation type at each set of input parameters were calculated by averaging over the last 30\% of the time trace.
Each plot in Fig.~\ref{error} shows the percent error in saturated heat flux, $(Q^{Closed}-Q^{Kinetic})/Q^{Kinetic}$, for all parameter combinations for each closure scheme. 
Versions of figures ~\ref{flux} and ~\ref{error} including $\nu=0$ can be found in Appendix~\ref{flux_appendix}.  Comparisons are complicated somewhat by occasional shifts in transport levels that occur unpredictably in time, which introduces a level of uncertainty that cannot be eliminated within the scope of this paper.  This is a manifestation of metastable states, recently elucidated in Ref.~\cite{christen2021bistable}.  
    
As expected, truncation performs poorly at low values of collisionality, but improves at high collisionality where the simulations become more fluid-like. Truncation still performs relatively poorly at high collisionality for low gradient drive, but performs well when both collisionality and gradient drive are large---i.e., for parameters at which other effects (gradient drive or collisions) dominate phase mixing.

The HP closure works well in the low collisionality regime for which it was designed (note the small errors at $\nu=0.01$).  However, its performance deteriorates as collisionality is increased. 

%\ASnote{Explanation for HP performance - not sure if the explanation commented out below still applies since HP no longer seems to deteriorate as gradient drive increases.}
%The HP errors are also larger as the gradient drive increases and the phase mixing physics must compete with $\omega_*$ physics and the nonlinearlity.\ASnote{Explain $\omega_*$ physics}  This suggests that the HP closure is well-suited for the its targeted regime (a regime where phase mixing dominates).  However, when other physics enters (strong gradient drive, nonlinearity, and/or collisions), it is too restrictive.

The HPC closure is designed to simultaneously include the effects of collisions and phase mixing in the appropriate limits.
As expected, it exhibits notable improvement over the HP closure in the high collisionality regime while also retaining the strong performance of the collisionless HP closure at low collisionality. This closure performs poorly only at intermediate levels of gradient drive ($\omega_T = 7-9$) and in one simulation at $\omega_T,\nu = 15,0.2$. This closure appears to be highly effective and its performance is only surpassed by the LMM closure trained at multiple parameter points, described below.

%but produces substantial errors at low collisionality. This is similar to the trend exhibited by simple truncation with one major difference---the Snyder closure is superior at low gradient drive and good performance extends to lower collisionality. 

The LMM-Middle closure based on the kinetic simulation at $\omega_T, \nu = 9, 0.1$ surprisingly performs poorly in the simulation at its training parameter point, likely due to the propensity of this system toward metastable states (note the sudden jump in the LMM time trace toward the end of the simulation).  The model does, however, perform well in nearby regions in the middle of our scanned parameter space. In fact, this closure extends throughout parameter space quite well, displaying low errors everywhere except in the top left corner (low collisionality and gradient drive). 

The LMM-Right closure based on the kinetic simulation at $\omega_T, \nu = 12,0.5$ produces very low errors at high collisionality and gradient drive, in the middle of the parameter region, and even at low gradient drive if collisionality is high. However, it also starts to deteriorate as simulations venture farther from the training point---at low collisionality and gradient drive.

%\DRHnote{need to discuss in referee responses the differences for LMM left from the previous version}\ASnote{Ok, will do. I think the best way will be to show a tile plot (in the referree response document) showing the saturation levels for the new and old SVD Left, so I'll get this done.}
The LMM-Left closure based on the kinetic simulation at $\omega_T, \nu = 6,0.01$ works well at low gradient drive regardless of collisionality. At high values of gradient drive however, its performance is poor. Specifically at the highest value of gradient drive, $\omega_T = 15$, this closure has extremely large errors. Inspecting Fig.~\ref{flux}, one can see the reason for this. The LMM-Left heat flux initially appears to saturate at an accurate level, but part way through the simulation, it jumps to a higher level and re-saturates. This behavior can also be seen in kinetic simulations at $\omega_T, \nu = 15, 0.2$ and $\omega_T, \nu = 15, 0.5$. When this jump occurs in the closed simulation but not the kinetic, it results in very large errors. These simulations illustrate the potential hazard of applying an LMM closure in a regime too far removed from its training point.  

We note several sets of simulations with sudden jumps between saturation levels, likely indicative of metastable states.  This is particularly clear in the lower right hand corner of the parameter grid, but can be observed throughout the parameter space.  It is likely that some fraction of the errors can be attributable to this phenomenon.  One example is the LMM-Middle training point, described above.  The large HPC error at $\omega_T=15,\nu=0.2$ is another candidate.  Unfortunately, this error can not be eliminated within the scope of this work. We do, however, probe this phenomenon in some more detail.

First, we ran a second set of kinetic simulations with identical physical parameters to the first set.  The initial condition is also identical save for a random component, which, in some cases can result in deviations in the long-time simulations.  Results are shown in Fig.~\ref{flux}.  We compared the heat flux saturation levels between the two sets of simulations to quantify the uncertainty in kinetic saturation levels. The percent difference between the saturation level in each initial simulation was compared to the saturation level of its counterpart with slightly different initial conditions. There was an 5\% RMS difference between the saturation levels in the two sets of simulations. There was one simulation at $\omega_T,\nu=7,0.5$ in which the simulation from the first set experienced a jump from one saturated state to another but the simulation from the second did not. So for this particular parameter point, there was a 20\% difference in the saturation levels. This illustrates the metastable nature of our system, which introduces some uncertainty into the determination of saturated heat flux levels.

\begin{figure}[H]
\includegraphics[width=0.5\textwidth]{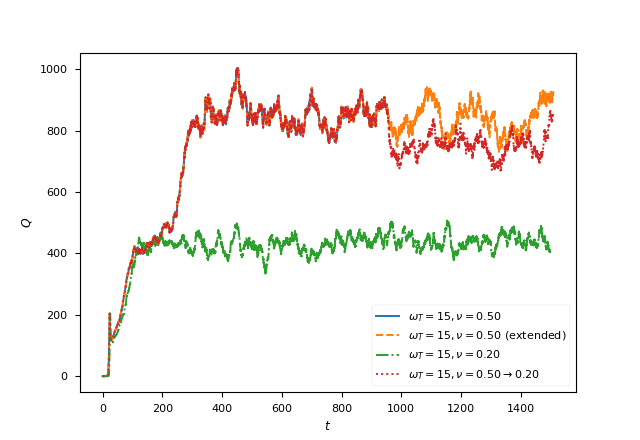}
\caption{
\label{collisionality_test}
}
\end{figure}

As another test, we probed the kinetic simulation at $\omega_T = 15,\nu=0.5$. 
 %which initially appeared to start saturating at a level similar to that of the $\omega_T = 15, \nu =0.2$ simulation but then jumped to a saturated state with higher $Q$. 
 We changed the collisionality of this simulation from $\nu=0.5$ to $\nu=0.2$ and restarted the simulation with the end-point of the previous simulation as the initial condition. The results of this test are shown in Fig.~\ref{collisionality_test}. The system finds a saturated state considerably higher than its counterpart with identical physical parameters.  The simulation which starts at $\nu=0.5$ initially saturates at $Q\approxeq 840$ but resaturates at $Q\approxeq 760$ once $\nu$ is switched to $\nu=0.2$. The simulation where $\nu=0.2$ the entire time is saturated at $Q\approxeq 460$. It appears that $Q=760$ and $Q=460$ are both stable states for the system with parameters $\omega_T,\nu = 15,0.2$. Which of these metastable states the system falls into depends on the previous state of the system. This test clearly demonstrates that our system has multiple metastable states and exhibits hysteresis. We refer the reader again to a recent, more detailed, study of this phenomenon~\cite{christen2021bistable, Heinonen}.

 %\textcolor{orange}{\ASnote{remove this paragraph if it is unnecessary. You mention in a previous paragraph that the jumps complicate comparisons} The existence of metastable states in our system certainly complicates comparisons with reduced models. More detailed investigation of these states will be left for future work. One interesting topic for further investigation would be measuring the ability of different closures to mirror kinetic transitions between metastable states and possibly developing a new version of the LMM closure that does exhibit this property.}
 
 %to see if it would drop back to its initial saturated level. The result was that it experienced a slight drop of ??\% but did not drop all the way back to a level similar to the $\omega_T=15,\nu=0.2$ simulation. Not sure what the implication of this second test is, maybe you can say?}

We note that the LMM-Right and LMM-Middle closures are similarly accurate to the HP and HPC closures throughout the parameter space, even taking into account the simulations far removed from the training point.
The LMM-left closure exhibits the worst performance of all the closures due to the large late-time jumps in saturation levels at high gradient drive.  

Ultimately, we envision the LMM closure applied in scenarios where kinetic training simulations can be supplied sparsely throughout parameter space.  Consequently, we show in Fig.~\ref{error} the closure called \textit{LMM-Optimal}, which is defined by selecting the LMM-closure that is trained at the nearest point in parameter space.  For example, at $\omega_T, \nu = 5, 0.05$,  the LMM-Left closure with coefficients extracted from the $\omega_T, \nu = 6,0.01$ kinetic simulation is used and at $\omega_T, \nu = 15,0.5$, the LMM-Right closure with coefficients extracted from the $\omega_T, \nu = 12,0.5$ kinetic simulation is used. 
In Fig.~\ref{error}, we include an LMM-Optimal closure that makes use of all three training points and another that uses only the left and right training points. Using two training points instead of three only slightly reduces performance; using three training points results in an RMS error of 9\% and using two yields an RMS error of 12\%.
Tables~\ref{opt_table} and~\ref{opt_table2} in Appendix~\ref{flux_appendix} show which training simulation was used to run the LMM-closed simulation at each grid point for the LMM-Optimal (3 Point) and LMM-Optimal (2 Point) plots in Figure~\ref{error}. Choosing the coefficients this way---using the LMM coefficients extracted from the nearest kinetic simulation to each grid point---results in excellent performance.  Only two parameter points exhibit errors above 20\% and the RMS error is only 9\% when three training points are used.   

The utility of the LMM closure would depend on efficient exploration of parameter space.  A database of closure information could be maintained as a parameter space is explored.  Expensive kinetic simulations could be performed sparsely, guided by a rigorous statistical framework like Bayesian optimization.  Fast fluid simulations, could then select the closure information at the nearest available parameter point.

%This procedure would be most powerful when carried out within a rigorous statistical framework like Bayesian optimization. %This would be a very computationally efficient and accurate way of conducting parameter scans. 

%\DRHnote{I think we don't need the following (commented out) paragraph--I think we've drawn the distinction clearly in early sections.}\ASnote{Ok, agreed}
%As noted earlier in Sec.~\ref{svd closure}, an HP-style closure could be formulated in a similar fashion to the LMM closure as a projection onto basis vectors. The fundamental difference between the HP closure and the LMM closure is the method used to choose those basis vectors - the HP style closure would use the eigenvectors of the linear system while the LMM closure uses basis vectors extracted from the nonlinear system. This would appear to be what allows the LMM closure to capture nonlinear effects in a way the HP closure cannot.\ASnote{Don't know if we want this paragraph or where it should go. I guess we want to draw a parallel between the LMM and HP closures somewhere in the paper?}

\section{Summary and Conclusions}
\label{summary}

%We have compared several closure methods for a relatively simple turbulent system---ITG/ETG driven turbulence in an unsheared slab---throughout the relevant 2D parameter space (collisionality and gradient drive).  Comparisons between four-moment fluid systems and a kinetic treatment demonstrate that simple truncation performs poorly, with errors roughly at the level of $20-50\%$, while the HP closure performs much better, particularly in the low collisionality regime.  Our new LMM closure outperforms both throughout the parameter space with errors generally less than $10\%$.  

We have introduced the LMM closure---a new method for formulating closures based on kinetic simulation data.  Here it is applied to the problem of phase mixing in a relatively simple turbulent system---gradient driven turbulence in an unsheared slab.  The LMM closure utilizes an optimal set of basis vectors from a nonlinear kinetic simulation to formulate closure coefficients in a fluid system.  We have compared this method to several closure methods throughout the relevant 2D parameter space (collisionality and gradient drive).  

First, we demonstrate that the LMM closure addresses the observation in several recent papers that phase mixing rates are substantially decreased in a turbulent system from the linear expectations.  The HP closure is observed to be too dissipative, over-estimating nonlinear values.  We demonstrate that the LMM closure reproduces phase mixing rates quite accurately. 
%lower phase mixing rates that are quite accurate for most parameter regimes although it does underestimate phase mixing rates for some parameter regimes - regimes in which the LMM-left closure is used.}  

Comparisons between heat fluxes for four-moment fluid systems and the kinetic system demonstrate that simple truncation performs rather poorly.  The HP and HPC closures perform much better than simple truncation.  In particular, the HPC closure appears to be highly effective, resulting in relatively low errors and exhibiting no systematic breakdown in parameters space.  The LMM closure is also highly effective when using multiple, sparse, training points, producing the lowest errors of all the closures.   

Our results suggest that the LMM closure has various advantages as well as drawbacks, which we summarize here.
Advantages:
\begin{itemize}
\item Capacity to capture the reduced phase mixing rates observed in turbulent systems.
\item Accuracy throughout parameter space when trained sparsely.
\item Generalizable, in principle, to more comprehensive systems (e.g., that described in Ref.~\cite{mandell_dorland_landreman_2018}) and other closure terms (e.g. curvature terms).
\item Applicable to scenarios where analytic closures have not been formulated or are inaccurate.
\end{itemize}

Drawbacks:
\begin{itemize}
    \item Requirement for kinetic training simulations.
    \item Uncertain extrapolation throughout parameter space.
\end{itemize}

We expect these drawbacks can be mitigated by statistical frameworks (e.g., Bayesian optimization) designed to track and optimally minimize uncertainties throughout a parameter space.  

Various generalizations can be imagined.  For example, suitable basis vectors could perhaps be formulated without the need for a nonlinear kinetic simulation by, e.g., taking inspiration from (multiple) linear eigenmodes or conducting a deeper study of nonlinear modifications to the linear eigenmodes.    

This work also represents one of the most thorough examinations of HP closures in comparison with nonlinear kinetic simulations.  The HP closures were observed to produce phase mixing rates far above the nonlinear kinetic values in this gradient-driven system.  Various improvements could be envisioned to mitigate this,for example, retaining some additional number of moments in the system.  Notably, despite the discrepancy in phase mixing rates, the heat flux predictions were much more accurate and competitive with the LMM closure.  The connection (or lack thereof) between phase mixing rates and heat fluxes is a compelling open question.        

The application of closures like the one proposed here to simulations of more comprehensive tokamak or stellarator systems may enable efficient exploration of fusion configurations with the ultimate goal of optimizing fusion performance.

\clearpage    
{\em Acknowledgements.--} This research used resources of the Texas Advanced Computing Center (TACC) at The University of Texas at Austin.  The authors would like to acknowledge Michael Nastac for useful discussions.  This work was supported by U.S. DOE Contract No. DE-FG02-04ER54742 and U.S. DOE Office of Fusion Energy Sciences Scientific Discovery through Advanced Computing (SciDAC) program under Award Number DE-SC0018429.     
\pagebreak 

\begin{appendices}
\section{HP Closure}
\label{hp_appendix}
As introduced in~\cite{hammet90} and derived formally in~\cite{Smith97}, the HP/Smith-style closure coefficients are chosen by matching the approximate response $R_{00}^a(\omega)$ to the exact response $R_{00}(\omega)$, which is given by
\begin{equation}
    %R_{00}(\omega) = -iZ(\omega+i\nu)
    R_{00}(\omega) = -iZ(\omega)
    \label{exact}
\end{equation}
where $Z$ is the plasma dispersion function.

The HP closure for the N$^{th}$ moment, $ f_N(\omega)$ takes the form $ f_N = \sum_{i=0}^{N-1} A_i  f_i(\omega)$. When the first $p$ closure coefficients, $A_i$, are set to zero, the approximate response in terms of the orthogonal polynomials, $P_i(\omega)$, and conjugate polynomials, $Q_i(\omega)$, becomes 
\begin{equation}
\mathrm{R}_{00}^{\mathrm{a}}(\omega)=i \frac{Q_{n}(\omega)-A_{n-1} Q_{n-1}(\omega)-\cdots-A_{p} Q_{p}(\omega)}{P_{n}(\omega)-A_{n-1} P_{n-1}(\omega)-\cdots-A_{p} P_{p}(\omega)}
\label{R00a}  
\end{equation}
and matches the exact response to order $(O(\omega^{n+1+p}))$.
The orthogonal and conjugate polynomials are expressed as
\begin{equation}
\begin{aligned}
P_{j}(\omega) &=P_{j, 0}+P_{j, 1} \omega+\cdots+P_{j, j} \omega^{j} \\
&=P_{j, 0}^{\omega_{0}}+P_{j, 1}^{\omega_{0}}\left(\omega-\omega_{0}\right)+\cdots+P_{j, j}\left(\omega-\omega_{0}\right)^{j}
\end{aligned}
\end{equation}
and
\begin{equation}
\begin{aligned}
Q_{j}(\omega) &=Q_{j, 0}+Q_{j, 1} \omega+\cdots+Q_{j, j} \omega^{j} \\
&=Q_{j, 0}^{\omega_{0}}+Q_{j, 1}^{\omega_{0}}\left(\omega-\omega_{0}\right)+\cdots+Q_{j, j}\left(\omega-\omega_{0}\right)^{j}
\end{aligned}
\end{equation}

This becomes the $[n,q,\omega_0=0]$ Pade approximant to the exact response for $q=n-p$ if the $q$ remaining coefficients are chosen to match the first $q$ terms of the Taylor series 
\begin{equation}
    R_{00}(\omega) = r_0 + r_1\omega + r_2\omega^2 +\cdots
    \label{R00 series}
\end{equation}
Since we are matching in the $\omega \rightarrow 0$ limit, we can substitute
\begin{equation}
Z(\zeta)=i \sqrt{\pi} e^{-\zeta^{2}}-2 \zeta\left(1-\frac{2 \zeta^2}{3}+\frac{4 \zeta^4}{15}-\frac{8 \zeta^6}{105}+\ldots\right),
\end{equation}
the Taylor expansion for the plasma dispersion function in the low frequency limit, into~\ref{exact} to define $r_i$ in~\ref{R00 series}.

Setting $R_{00}^a(\omega)=R_{00}(\omega)$ results in the following equation
\begin{equation}
\begin{aligned}
&\left(r_{0}+\ldots+r_{q-1} \omega^{q-1}\right)(A_{n-1} P_{n-1}(\omega)\\
&+\ldots+A_{p} P_{p}(\omega))\\
&-i\left(A_{n-1} Q_{n-1}(\omega)+\ldots+A_{p} Q_{p}(\omega)\right)\\
&=\left(r_{0}+\ldots+r_{q-1} \omega^{q-1}\right) P_{n}(\omega)-i Q_{n}(\omega)
\end{aligned}
\end{equation}
Choosing the coefficients ${A_{n-1},\dots,A_p}$ so that the coefficients of the different powers of $\omega$ on the RHS and LHS match results in the closure.

This matching can be done by solving the following matrix equation
\small
\begin{equation}
    \begin{array}{l}
\Bigg(\begin{bmatrix}
r_{0} & 0 & \ldots & 0 \\
r_{1} & r_{0} & \ldots & 0 \\
\vdots & \vdots & \ddots & \vdots \\
r_{q-1} & r_{q-2} & \ldots & r_{0}
\end{bmatrix}\\
\begin{bmatrix}
P_{n-1,0} & P_{n-2,0} & \ldots & P_{p, 0} \\
P_{n-1,1} & P_{n-2,1} & \ldots & P_{p, 1} \\
\vdots & \vdots & \ddots & \vdots \\
P_{n-1, q-1} & P_{n-2, q-1} & \ldots & P_{p, q-1}
\end{bmatrix}\\
-i\begin{bmatrix}
Q_{n-1,0} & Q_{n-2,0} & \ldots & Q_{p, 0} \\
Q_{n-1,1} & Q_{n-2,1} & \ldots & Q_{p, 1} \\
\vdots & \vdots & \ddots & \vdots \\
Q_{n-1, q-1} & Q_{n-2, q-1} & \ldots & Q_{p, q-1}
\end{bmatrix}\Bigg)
\times\begin{bmatrix}
    A_{n-1}\\
    A_{n-2}\\
    \vdots\\
    A_{p}
\end{bmatrix}\\
=\begin{bmatrix}
r_{0} & 0 & \ldots & 0 \\
r_{1} & r_{0} & \ldots & 0 \\
\vdots & \vdots & \ddots & \vdots \\
r_{q-1} & r_{q-2} & \ldots & r_{0}
\end{bmatrix}
\begin{bmatrix}
    P_{n,0}\\
    P_{n,1}\\
    \vdots\\
    P_{n,q-1}
\end{bmatrix}
-i\begin{bmatrix}
    Q_{n,0}\\
    Q_{n,1}\\
    \vdots\\
    Q_{n,q-1}
\end{bmatrix}
\end{array}
\label{matrix solution}
\end{equation}
\normalsize
for the closure coefficients ${A_n,\dots,A_p}$.

We use $P_n(x) = H_n(x)$ defined by
\begin{equation}
    H_n(x) = \frac{(-1)^ne^{x^2}}{\sqrt{2^n n! \sqrt\pi}}
    \frac{d^n}{dx^n}e^{-x^2}
\end{equation}
satisfying the recurrence relation
\begin{equation}
    H_n(x) = \sqrt{\frac{2}{n}} x H_{n-1}(x) - \sqrt{\frac{n-1}{n}}H_{n-2}(x)
    \label{recurrence}
\end{equation}
and conjugate polynomials $Q_n(x)$ constructed by requiring that they satisfy the same recurrence relation as our Hermite polynomials and the conditions
\begin{equation}
    Q_0(x) = 0,\, Q_1(x) = \pi^{-1/4}\sqrt{2}.
    \label{conjugate_initial}
\end{equation}

The orthogonal polynomial moments, our Hermite moments, are defined with respect to the Hermite polynomials as 
\begin{equation}
     f_n = \int_{-\infty}^{\infty} f(x) H_n(x) dx .
\end{equation}

These are the Hermite polynomials and Hermite decomposition used in our model as described in Sec.~\ref{reduced_hermite}.

\subsection{Verification of HP Coefficients}
\begin{figure}[ht]
    \centering
    \includegraphics[width=.5\textwidth]{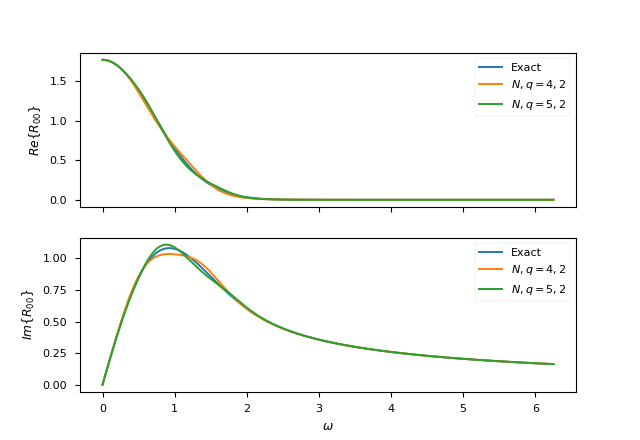}
    \caption{Real and imaginary parts of the exact  response function and the approximate response functions for the N,q=4,2 and N,q=5,2 HP closures.}
    \label{fig:R00 plot}
\end{figure}

The HP closure that we focus on closes at the $4^{th}$ moment with the N,q = 4,2 HP closure. For this closure we have $A_3 = -1.759 i,\, A_2 = 0.755$.
This gives the following HP closure for our system in DNA
\begin{equation}
    f_{\mathbf k,4} = -1.759i*sgn(k_z)f_{\mathbf k,3}+0.755f_{\mathbf k,2}
\label{f4_collisionlessclosure}
\end{equation}
The $sgn(kz)$ is present in the $f_{\mathbf k,3}$ term to ensure damping for both positive and negative $k_z$.  As shown in Fig.~\ref{fig:R00 plot}, this closure produces an excellent match to the response function.

To construct the collisional HP closure we instead need the coefficients for a closure at the $5^{th}$ moment. We use the N,q = 5,2 closure which takes the form $f_{\mathbf k,5} = A_4sgn(k_z)f_{\mathbf k,4} + A_3f_{\mathbf k,3}$ and has coefficients $A_4 = - 1.805i$ and $A_3 = 0.801$. 

To verify the correctness of our choice of coefficients for the $N,q = 4,2$ and $N,q = 5,2$ HP closures, we plot the approximate response functions produced by these closures along with the exact response function in Fig.~\ref{fig:R00 plot}. Both closures show good agreement.

\section{The HPC Closure} 
\label{snyder_appendix}
The HP closure is designed for a collisionless regime.  However, there is a method for modifying the HP closure to include the effects of collisions developed in~\cite{Snyder97}. 

In order to derive the collisional closure for our system, we start by using Smith's method to determine the optimal collisionless closure for $f_{\mathbf k,5}$ with N,q=5,2 as described above. This yields the collisionless Smith/HP style closure for the 5$^{th}$ moment:
\begin{equation}
    f_{\mathbf k,5}=sgn(k_z)A_4 f_{\mathbf k,4}+A_3 f_{\mathbf k,3}
    \label{f5_collisionless}
\end{equation}
with coefficients $A_4 = -1.805i$ and $A_3 = 0.801$. Notice the $sgn(kz)$ in the $f_{\mathbf k,4}$ term. This is included to ensure damping at both positive and negative $k_z$.

We then plug this expression for $f_{\mathbf k,5}$ into the linearized time evolution equation for $f_{\mathbf k,4}$ including collisions which can be obtained by taking $n=4$ in Eq.~\ref{dnaeq} and excluding the nonlinear term. 
This gives our first expression for $\partial f_4/\partial t$.

\begin{multline}
\pdv{f_4}{t} = -4\nu f_{\mathbf k,4} - \sqrt5ik_zf_{\mathbf k,5} - 2ik_zf_3\\
\pdv{f_4}{t} = -4\nu f_{\mathbf k,4} - \sqrt5ik_z(sgn(k_z)A_4 f_{\mathbf k,4}+A_3 f_{\mathbf k,3} ) - 2ik_zf_3
\label{df4dteq1}    
\end{multline}

We then take the low frequency limit, $\omega \rightarrow 0$, of Eq.~\ref{df4dteq1}($\partial f_{\mathbf k,4}/\partial t = 0$), 

\begin{equation}
0 = -4\nu f_{\mathbf k,4} - \sqrt5ik_z(sgn(k_z)A_4 f_{\mathbf k,4}+A_3 f_{\mathbf k,3} ) - 2ik_zf_3
\label{0df4dteq1}    
\end{equation}
and solve the resulting equation for $f_{\mathbf k, 4}$ in terms of $f_{\mathbf k,3}$.
\begin{equation}
f_{\mathbf k,4} = \frac{-(2 + \sqrt5A_3)ik_zf_{\mathbf k,3}}{4\nu + \sqrt5 A_4sgn(k_z) ik_z}
\label{f4lfcoll}
\end{equation}
This expression for $f_{\mathbf k,4}$ is first order accurate in $\omega$ for small $\omega$.

%We seek a an expression for $\partial f_{\mathbf k,4}/\partial t$ that we can equate to Eq.~\ref{df4dteq1}, 
We seek an expression for $f_{\mathbf k,4}$ that is second order accurate in $\omega$, so we take the time derivative of both sides of Eq.~\ref{f4lfcoll}.

\begin{equation}
\pdv{f_{\mathbf k,4}}{t} = \frac{-(2 + \sqrt5A_3)ik_z\partial f_{\mathbf k,3}/\partial t}{4\nu + \sqrt5 A_4sgn(k_z) ik_z}
\end{equation}

Substituting in $\partial f_{\mathbf k,3}/\partial t$, which is obtained from Eq.~\ref{dnaeq}, we get our second expression for $\partial f_{\mathbf k,4}/\partial t$.
\begin{equation}
\pdv{f_{\mathbf k,4}}{t} = \frac{\splitfrac{-(2 + \sqrt5A_3)ik_z (-ik_z\sqrt3f_{\mathbf k,2} }{ -2ik_zf_{\mathbf k,4} - 3\nu f_{\mathbf k,3}) }}{4\nu + \sqrt5 A_4sgn(k_z) ik_z}
\label{df4dteq2}
\end{equation}

Taking the low frequency limit again, we have
\begin{equation}
0 = \frac{ \splitfrac{-(2 + \sqrt5A_3)ik_z (-ik_z\sqrt3f_{\mathbf k,2}}{ -2ik_zf_{\mathbf k,4} - 3\nu f_{\mathbf k,3})} }{4\nu + \sqrt5 A_4sgn(k_z) ik_z}
\label{0df4dteq2}
\end{equation}

We then equate the right hand sides of Eq.~\ref{0df4dteq1} and Eq.~\ref{0df4dteq2} and solve for $f_{\mathbf k,4}$ to obtain the collisional N,q=4,2 closure.
\small
\begin{equation}
f_{\mathbf k,4} = \frac{\splitfrac{(-(7\sqrt5iA_3 + 14i)k_z\nu + (5A_3 + 2\sqrt5)}{A_4sgn(k_z)k_z^2)f_{\mathbf k,3}  +(k_z^2 ( 15A_3 + 2\sqrt3))f_{\mathbf k,2}} }{\splitfrac{16\nu^2 + 8\sqrt5 ik_z\nu A_4sgn(k_z)}{ - k_z^2(5A_4^2 + 2\sqrt5 A_3+4) }}
\label{f4_collclosure_symbolic}
\end{equation}
\normalsize
%\begin{multline}
%f_{\mathbf k,4} = \frac{-(7\sqrt5iA_3 + 14i)k_z\nu + (5A_3 + 2\sqrt5)A_4sgn(k_z)k_z^2}{16\nu^2 + 8\sqrt5 ik_z\nu A_4sgn(k_z) - k_z^2(5A_4^2 + 2\sqrt5 A_3+4) }f_{\mathbf k,3} \\
%+ \frac{k_z^2 ( 15A_3 + 2\sqrt3) }{16\nu^2 + 8\sqrt5 ik_z\nu A_4sgn(k_z) - k_z^2(5A_4^2 + 2\sqrt5 A_3+4) }f_{\mathbf k,2}
%\label{f4_collclosure_symbolic}
%\end{multline}

Plugging in $A_4 = -1.805i$ and $A_3 = 0.801$ which we know from using Smith's method to determine the collisionless 5,2 closure, we get the final numerical expression for our 4,2 collisional closure:
\begin{multline}
f_{\mathbf k,4} = \frac{-3.051ik_z\nu - 1.759ik_z^2sgn(k_z)}{1.838\nu^2 + 3.709 k_z\nu sgn(k_z) + k_z^2}f_{\mathbf k,3}\\
+ \frac{0.755 k_z^2}{1.838\nu^2 + 3.709 k_z\nu sgn(k_z) + k_z^2}f_{\mathbf k,2}
\label{f4_collclosure}
\end{multline}
This is the HPC closure, a 2$^{nd}$ order accurate (for small $\omega$) closure for $f_{\mathbf k, 4}$ in terms of $f_{\mathbf k, 3}$ and $f_{\mathbf k,2}$ including collisional effects.

Note that if one takes the collisionless limit, $\nu\rightarrow0$, of Eq.~\ref{f4_collclosure}, the collisionless closure given in Eq.~\ref{f4_collisionlessclosure} is recovered.

\section{Hermite and Fluid Moments}
\label{hermite_appendix}
In a fluid or gyrofluid model, the $n^{th}$ moment is calculated as $\int v^n f dv$ where f is the kinetic or gyrokinetic distribution function respectively. The first four moments are physically meaningful: $n$ is density, $u$ is mean velocity, $p$ is pressure, and $q$ is heat flux. Moments beyond $q$ are not physically meaningful but may need to be calculated to ensure the accuracy of the model. 

The Hermite moments in our system are calculated as $ f_n = \int f(v) H_n(v) dv$ where $f(v)$ is the gyrokinetic distribution function. Since the Hermite polynomials contain only powers of $v$, the fluid moments are simply linear combinations of the Hermite moments. The relationship between the first 5 fluid moments and the first 5 Hermite moments is shown below.
\begin{align}
    n &= \int dv f(v) = \pi^{1/4} f_0\\
    u &= \int dv f(v) v = \frac{\pi^{1/4}}{\sqrt 2} f_1\\
    p &= \int dv f(v) v^2 = \frac{\pi^{1/4}}{\sqrt 2}  f_2 + \frac{\pi^{1/4}}{2} f_0\\
    q &= \int dv f(v) v^3 = \frac{\sqrt 3\pi^{1/4}}{2} f_3 + \frac{3\pi^{1/4}}{2\sqrt 2} f_1\\
    r &= \int dv f(v) v^4 = \frac{\sqrt3\pi^{1/4}}{\sqrt2} f_4 + \frac{3\pi^{1/4}}{\sqrt2} f_2 + \frac{3\pi^{1/4}}{4} f_0 
\end{align}
%\begin{landscape}
\section{Additional Heat Flux plots}
\label{flux_appendix}
\begin{figure*}[ht]
\centering
    \includegraphics[width=1\textwidth]{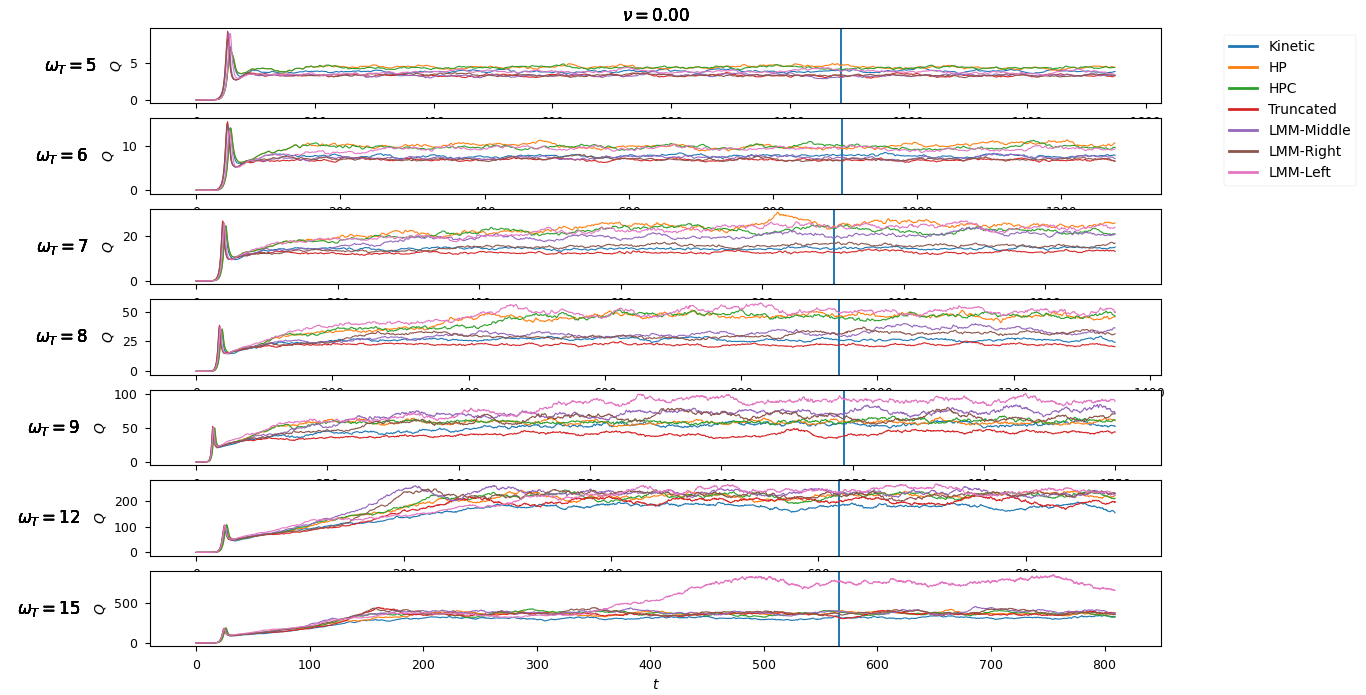}
    \captionsetup{width=\textwidth}
    \caption{
    \label{all_flux}
        Time traces of the total radial heat flux ($Q$) for Kinetic (blue), HP (orange), HPC (green), truncated (red), LMM-Middle (purple), LMM-Right (brown), and LMM-Left (pink)  simulations for temperature gradient drives ($\omega_T$) ranging from 5 to 15 (increasing downward by panel) and collision frequency $\nu = 0$. The vertical blue lines show the cutoff point - 70\% of the simulation time - after which each heat flux curve is averaged to get the final saturation level. 
        }
\end{figure*}
%\end{landscape}
\begin{figure*}[ht]
\centering
\includegraphics[width=\textwidth]{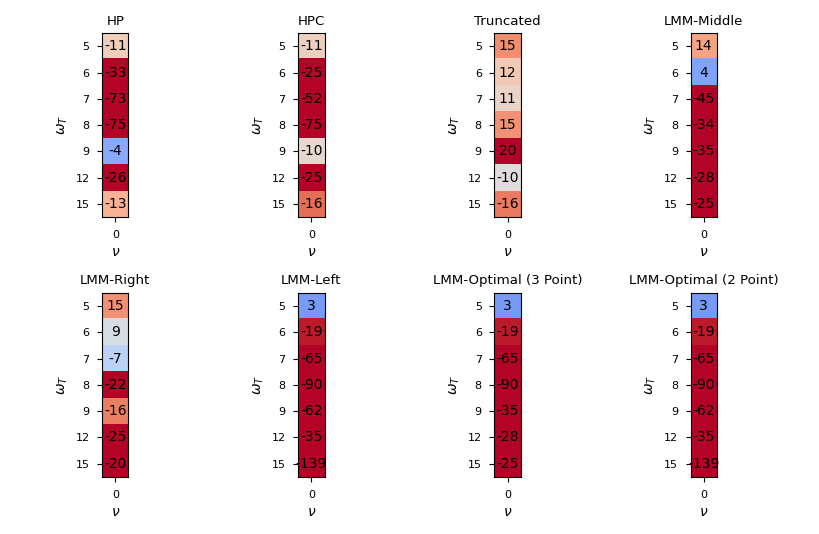}
\caption{
\label{all_error}
Percent error in saturated heat flux for each closure (HP, HPC, Truncation, LMM-Middle, LMM-Right, LMM-Left, LMM-Optimal (3 Point), LMM-Optimal (2 Point)) as compared against the kinetic simulation for temperature gradient drives ($\omega_T$) ranging from 5 to 15 (increasing downward) and collision frequency $\nu = 0$. Percent errors are calculated as $(Q^{Closed} - Q^{Kinetic})/Q^{Kinetic} \times 100$ where $Q^{closed}$ is calculated by averaging the last 30\% of the time trace of the heat flux from the closed simulation and $Q^{Kinetic}$ is calculated by averaging the last 30\% of the time trace of the heat flux from the kinetic simulation. Average errors are calculated by averaging the absolute value of the errors.}
\end{figure*}

\begin{table*}[ht]
    \centering
    \begin{tabular}{|c|c|c|c|c|c|c|}
       \hline
       \backslashbox{$\omega_T$}{$\nu$}  & 0 & 0.01 & 0.05 & 0.1 & 0.2 &0.5\\
       \hline
       5  & Left & Left  & Left & Left & Left & Left\\
       \hline
       6  & Left & Left & Left & Left & Left & Middle\\
       \hline
       7  &  Left & Left & Left & Middle & Middle & Middle\\
       \hline
       8  & Left & Left & Middle & Middle & Middle & Right\\
       \hline
       9  & Middle & Middle & Middle & Middle & Middle & Right\\
       \hline
       12 & Middle & Middle & Middle & Middle & Right & Right\\
       \hline
       15 & Middle & Middle & Middle & Middle & Right & Right\\
       \hline
    \end{tabular}
    \caption{
    \label{opt_table}
    This table shows which training simulation, equivalently which set of LMM-closure coefficients, was used to produce the LMM-Optimal (3 Point) error plot in Figures~\ref{error} and~\ref{all_error}.
    At each grid point in parameter space, the set of coefficient from the training simulation that lies closest to that grid point is used.
    For example, at $\omega_T, \nu = 5, 0.05$,  the LMM-Left closure with coefficients extracted from the $\omega_T, \nu = 6,0.01$ kinetic simulation is used and at $\omega_T, \nu = 15,0.5$, the LMM-Right closure with coefficients extracted from the $\omega_T, nu = 12,0.5$ kinetic simulation is used.
    }
    \label{tab:my_label}
\end{table*}

\begin{table*}[ht]
    \centering
    \begin{tabular}{|c|c|c|c|c|c|c|}
       \hline
       \backslashbox{$\omega_T$}{$\nu$}  & 0 & 0.01 & 0.05 & 0.1 & 0.2 &0.5\\
       \hline
       5  & Left & Left  & Left & Left & Left & Left\\
       \hline
       6  & Left & Left & Left & Left & Left & Left\\
       \hline
       7  &  Left & Left & Left & Left & Left & Right\\
       \hline
       8  & Left & Left & Left & Left & Right & Right\\
       \hline
       9  & Left & Left & Left & Right & Right & Right\\
       \hline
       12 & Left & Left & Right & Right & Right & Right\\
       \hline
       15 & Left & Right & Right & Right & Right & Right\\
       \hline
    \end{tabular}
    \caption{
    \label{opt_table2}
    This table shows which training simulation, equivalently which set of LMM-closure coefficients, was used to produce the LMM-Optimal (2 Point) error plot in Figures~\ref{error} and~\ref{all_error}.
    At each grid point in parameter space, the set of coefficient from the training simulation that lies closest to that grid point is used.
    For example, at $\omega_T, \nu = 5, 0.05$,  the LMM-Left closure with coefficients extracted from the $\omega_T, \nu = 6,0.01$ kinetic simulation is used and at $\omega_T, \nu = 15,0.5$, the LMM-Right closure with coefficients extracted from the $\omega_T, nu = 12,0.5$ kinetic simulation is used.
    }
    %\label{tab:my_label}
\end{table*}

\clearpage
\end{appendices}

\clearpage
\bibliographystyle{ieeetr}
\bibliography{arxiv} {}
    
\end{document}